\documentclass[12pt]{iopart}

\usepackage{iopams}
\usepackage{graphics}
\usepackage{graphicx}
\usepackage{dcolumn}
\usepackage{epsfig}
\usepackage{bm}
\usepackage{color}

\usepackage{amssymb}
\usepackage{lineno}
\usepackage{epsfig}
\usepackage{subfigure}
\begin{document}

\title[]{Predicted superconductivity of Ni$_2$VAl and pressure dependence of superconductivity in Ni$_2$NbX (X = Al, Ga and Sn) and Ni$_2$VAl}

\author{P. V. Sreenivasa Reddy and V. Kanchana$^*$}

\address{Department of Physics, Indian Institute of Technology Hyderabad, Kandi, Medak-502 285, Telangana, India.}

\author{G. Vaitheeswaran}
\address{Advanced Centre of Research in High Energy Materials (ACRHEM), University of Hyderabad, Prof. C. R. Rao Road, Gachibowli, Hyderabad 500 046, Telangana, India.}

\author{David J. Singh}
\address{Department of Physics and Astronomy, University of Missouri-Columbia, Columbia, Missouri-65211, USA.}

\ead{kanchana@iith.ac.in}
\begin{abstract}
A first-principles study of the electronic and superconducting properties of the Ni$_2$VAl Heusler compound is presented. The electron-phonon coupling constant of $\lambda_{ep}$ = 0.68 is obtained, which leads to a superconducting transition temperature of T$_c$ = $\sim$4 $K$ (assuming a Coulomb pseudopotential $\mu^*$ = 0.13), which is a relatively high transition temperature for Ni based Heusler alloys. The electronic density of states reveals a significant hybridization between Ni-$eg$ and V-$t_{2g}$ states around the Fermi level. The Fermi surface, consisting of two electron pockets around the X-points of the Brillouin zone, exhibits nesting and leads to a Kohn anomaly of the phonon dispersion relation for the transverse acoustic mode TA2 along the (1,1,0) direction, which is furthermore found to soften with pressure. As a consequence, T$_c$ and $\lambda_{ep}$ vary non-monotonically under pressure. The calculations are compared to similar calculations performed for the Ni$_2$NbX (X = Al, Ga and Sn) Heusler alloys, which experimentally have been identified as superconductors. The experimental trend in T$_c$ is well reproduced, and reasonable quantitative agreement is obtained. The calculated T$_c$ of Ni$_2$VAl is larger than either calculated and observed T$_c$s of any of the Nb compounds. The Fermi surfaces of Ni$_2$NbAl and Ni$_2$NbGa consist of only a single electron pocket around the  X point, however under compression second electron pocket similar to that of Ni2VAl emerges only in Ni2NbAl and the Tc increases non monotonically in all the compounds. Fermi surface nesting and associated Kohn anomalies are a common feature of all four compounds, albeit weakest in Ni$_2$VAl.

\end{abstract}

\vspace{2pc}

\maketitle

\section{Introduction}
Heusler alloys are ternary intermetallic compounds of the form X$_2$YZ, where X is generally a transition metal, Y is yet another transition metal from group VIIIB-IB and Z is a $sp$ metal or a metalloid. These compounds display a wide range of  physical properties including halfmetallicity\cite{de Groot,Kandpal,Galehgirian,Feng}, magnetic ordering\cite{Gofryk,Kaczorowski}, heavy fermion behaviour\cite{Gofryk1,Nakamura,Takayanagi}, shape memory effect \cite{ Wuttig, Sutou} and thermoelectricity\cite{Tobola,Bhattacharya}. Some are superconductors having a  superconducting transition temperature (T$_c$) ranging from 0.74 K (HfNi$_2$Al) to 4.7 K (YPd$_2$Sn). Superconductivity was first reported for the Heusler alloys by Ishikawa et al. \cite{Ishikawa}, where they focused mainly on the systems RPd$_2$Sn and RPd$_2$Pb with R being a rare-earth metal. Among the known Heusler superconductors, Ni-based alloys Ni$_2$NbX (X = Al, Ga and Sn) have attracted much attention due to their intermediate electron phonon coupling constant \cite{S Waki,Boff,JH Wernick}. Interestingly, the Ni$_2$NbX (X = Al, Ga and Sn) compounds are paramegnets and are found to be superconductors although Ni is a ferromagnet and Ni compounds are often magnetic. Here we predict Ni$_2$VAl to be a superconductor. Importantly, this provides an experimentally testable prediction that if confirmed and taken in conjunction with the correct predictions for the related compounds, would strongly restrict the possible role of spin-fluctuations associated with Ni magnetism in the superconducting properties of these phases. 

In these compounds the presence of Nb works against magnetism associated with Ni leading to the paramagnetic ground states of these compounds \cite{S Waki, JH Wernick}. The same paramagnetic nature is recently observed in ScT$_2$Al (T= Ni, Pd, Pt, Cu, Ag, Au)\cite{Bennorf}. The Ni$_2$NbX superconductors have T$_c$ of 2.15 K (Ni$_2$NbAl)\cite{S Waki}, 1.54 K (Ni$_2$NbGa)\cite{S Waki} and 2.9 K (ref.15) (3.4 K (ref.16,17)) (Ni$_2$NbSn) with calculated electron phonon coupling constants\cite{S Waki} ($\lambda_{ep}$) of 0.52, 0.50 and 0.61 respectively. Electronic structures and cohesive properties of Ni$_2$NbAl and Ni$_2$VAl were studied by Lin et al.\cite{Lin} and a large value of cohesive energy is observed with a pronounced hybridization between Ni-Nb/V-Al atoms. The interaction between these atoms creates deep valleys in density of states which separate bonding and anti-bonding region. It is this type of covalency that works against magnetism.

In prior work \cite{PVSR}, we have  studied electronic structure of Ni$_2$NbAl and Ni$_2$VAl and superconductivity of Ni$_2$NbAl both at ambient as well as under compression and observed a change in Fermi surface (FS) topology at a pressure of around 17 $GPa$ in Ni$_2$NbAl. The change in the FS topology leads to the non-monotonic variation in the superconducting transition temperature. In case of Ni$_2$VAl we did not find any FS topology changes in the pressure range studied. Here we focus in detail on the superconductivity of Ni$_2$VAl and report results on Ni$_2$NbGa and Ni$_2$NbSn. No other reports are available for Ni$_2$NbGa and Ni$_2$NbSn regarding the electronic structure, elastic and vibrational properties. 

 Superconductivity in conventional intermetallics has been traditionally discussed in terms of the density of states at the Fermi level and the number of valence electrons per atom. From the Matthias rule\cite{Matthias, Matthias1}, the number of valence electrons per atom should be close to 5 or 7. Even though the Heusler superconducting compounds follow the above prescription, the superconducting transition temperatures of these compounds are relatively low. Among the compositional elements of the Ni$_2$VAl, vanadium is reported to be a superconductor with superconducting transition temperature ($T_c$) of 5.3 $K$\cite {Ishizuka} and 3.6 $K$\cite{Vaitheeswaran}.
From previous literatures \cite{Vaitheeswaran}, the total density of states (DOS) value of bcc Vanadium is 1.46 (states/eV/f.u.). Pressure effect on the T$_c$ has been studied experimentally and theoretically for bcc Vanadium\cite {Ishizuka,Vaitheeswaran,Suzuki,Suzuki1} upto 120 $GPa$ in which the T$_c$ is found to increase linearly with pressure. However, high density of states also favours spin-fluctuations, which work against electron-phonon superconductivity\cite{Berk}. 
From our previous work the total DOS of Ni$_2$VAl is observed to be more than the value of vanadium and the number of electrons/atom is observed to be 7 in the present compound obeying the  Matthias rule. 

We also studied the pressure dependence, as pressure provides a tuning parameter that is very useful in understanding trends and mechanisms. Previous studies on Hf based Heusler alloys have shown an increase in the superconducting transition temperature with decreasing lattice parameter\cite{Klimczuk}, while certain other cases show an increase in T$_c$ with increase in lattice constant\cite{Klimczuk}. We note that the behaviour can also be more complex under pressure \cite{Singh}. Fundamentally, superconductivity is an instability of the Fermi surface, and so pressure induced changes in Fermi surface can lead to non-trivial changes in superconducting properties.
 
Another interesting feature that connects with superconductivity is the presence of van Hove singularities\cite{Vanhove}, in the electronic structure, leading to peaks in the density of states as found in some of the Pd based Huesler compounds \cite{Winterlik1,Ramesh Kumar,Ming}. Interesting behaviour might be anticipated if the Fermi level can be brought to this peak by means of alloying. Another noteworthy point present in these compounds is the softening of the TA2 acoustic phonon modes, which can be well correlated with the FS nesting and the corresponding nesting vector decides the position of the Kohn anomaly present in these compounds. Further to these, we also find a softening in the acoustic phonons under pressure leading to the variation in the T$_c$, which is discussed in detail.

	The organization of the paper is as follows. Computational details are presented in the section 2. Results and discussions of ground state, electronic structure, mechanical, vibrational, superconducting properties and pressure effect on these properties are presented in the section 3. The conclusions are given in section 4.

\section{Computational details}
The Full Potential Linearized Augmented Plane Wave (FP-LAPW) method as implemented in Wien2k code \cite{Blaha} is used to calculate the ground state and electronic structure of the present compounds. We adopt Generalized Gradient Approximation (GGA) of Perdew-Burke-Ernzerhof (PBE)\cite{JPPE}. The wave functions expanded up to angular momentum $l$ = 10 inside the muffen-tin spheres. The radii of muffin tin spheres R$_{MT}$ were 1.78 $a.u$ for Ni, Nb and V, 1.67 $a.u$ for Al, 2.0 $a.u$ for Ga and 2.3 $a.u$ for Sn atoms. The plane wave cut off energy is used K$_{max}$ = 9/R$_{MT}$, where R$_{MT}$ is the smallest radius of muffin tin and K$_{max}$ is the magnitude of largest plane wave expansion. All the electronic structure calculations are performed with $44\times 44\times 44$ k-mesh in the Monkhorst-Pack \cite{Monkhorst} scheme which gives 2168 $k$-points in the irreducible part of the Brillouin Zone (BZ). Tetrahedron method \cite{tetra} was used to integrate the Brillouin zone. Energy convergence up to 10$^{-5}$ $Ry$ is used to get the proper convergence of the self consistent calculation. Birch-Murnaghan \cite{FBIR} equation of state is used to find the equilibrium lattice parameter and bulk modulus by fitting the total energies as a function of primitive cell volume. We have not found any significant change in the electronic structure at the Fermi level with the inclusion of spin-orbit coupling (SOC). So, the reported calculations are without SOC.

Phonon dispersions and electron-phonon interaction calculations were performed using the plane wave pseudopotential method (PWSCF) which is implemented in QUANTUM ESPRESSO \cite{Giannozzi}code. The GGA-PBE exchange correlation functional is used in the present calculations for all the compounds. The electron ion interaction is described by using norm-conserving pseudopotentials. The maximum plane wave cut-off energy (ecutwfc) was 90 $Ry$ and the electronic charge density was expanded up to 360 $Ry$ (In case of Ni$_2$VAl it is 140 and 560 $Ry$). A $16\times 16\times 16$ k-points grid within the BZ is used for the phonon calculations. Gaussian broadening of 0.02 $Ry$ and a $4\times 4\times 4$ uniform grid of $q$-points are used for phonon calculations.

\section{Results and discussions}
\subsection{Ground state, electronic structure properties}
	
The ground state properties of Ni$_2$NbX (X = Al, Ga and Sn) and Ni$_2$VAl are evaluated using the experimental lattice parameter and atomic positions\cite{S Waki, Boff, JH Wernick,Rocha,Villars}. The calculated equilibrium lattice constant and bulk modulus values are presented in Table-I along with the available experimental and other theoretical work and a good agreement is seen between the present values and earlier reports. The calculated bulk moduli of Ni$_2$NbAl, Ni$_2$NbGa and Ni$_2$VAl compounds are nearly same and found to be slightly high than Ni$_2$NbSn.

	We have calculated the band structure along the high symmetry directions in the irreducible Brillouin zone with and without inclusion of SOC. From Fig.1 (given for only Ni$_2$NbSn) it is seen that the SOC effect is very small around the Fermi level and we have proceeded with the further calculations excluding SOC. The band structures for all the compounds are given in Fig.2. The overall band shapes are similar for all the compounds studied but the Fermiology is different as seen from the band crossing at E$_F$. For all the compounds the lowest lying valence band arises mainly from the $s$-states of X (X = Al, Ga, Sn) atom. The bands near the E$_F$ are due to the hybridization of both Ni and Nb/V atoms. The band which cross the E$_F$ at X point in both Ni$_2$NbAl and Ni$_2$NbGa is due to the $d_{eg}$ states of Ni. The same band in Ni$_2$NbSn is crossing the E$_F$ at both X and W points. In addition to that we have an extra band at the X point in Ni$_2$NbSn which has Nb $d$ character. In the case of Ni$_2$VAl we have two bands to cross the E$_F$ at X point.

	The electronic density of states (DOS) is shown in Fig.3 along with the atom projected DOS. Even though these compounds are composed of different elements from different rows in the periodic table, the total DOS for all the compounds looks similar reflecting the similar band shapes. For all the compounds we observe valleys at energies of -6 $eV$, -1 $eV$, 0.5 $eV$. In the case of Ni$_2$NbSn there is another valley at around -3 eV. This feature indicates that the interaction between the constituent atom is strong \cite{Lin}. Among all the compounds Ni$_2$VAl has the highest value of DOS at E$_F$ with 3.51 $states/eV$ and remaining Ni$_2$NbX compounds have 2.27, 2.20 and 2.34 $states/eV$ for X = Al, Ga and Sn respectively. From the atom projected DOS we have observed that the primary contribution to the total DOS at E$_F$ is due to Ni atom ($d_{eg}$ states), the secondary contribution is due to Nb/V atom ($d_{t_{2g}}$ states) and the least contribution arises from X atom ($p$ states) (X = Al, Ga and Sn). The calculated Sommerfield coefficient $\gamma$ is also given in the Table-I along with the experimental values. 
There is a decreasing trend in N(E$_F$) from Ni$_2$VAl $\rightarrow$ Ni$_2$NbSn $\rightarrow$ Ni$_2$NbAl $\rightarrow$  Ni$_2$NbGa.  
Experimental specific heat can be calculated with $(1+\lambda)\times \gamma_{th}$. The $\lambda$ in the enhancement factor is related to but not identical to the superconducting $\lambda$ and includes in addition contributions from spin fluctuations and other interactions if present. The inferred values are in the range $\sim$0.5-1 for Ni$_2$NbAl and Ni$_2$VAl, in reasonable accord with the calculated superconducting $\lambda$ below. The values for Ni$_2$NbGa and especially Ni$_2$NbSn are anomalously low. The origin of this is not clear, and warrants further investigation. Site disorder in samples is one possibility. In any case, we also note that the N(E$_F$) are not high enough to place any of the compounds near Stoner criterion for ferromagnetism.

The van Hove singularity is observed in both valence and conduction bands at the L-point close to E$_F$ around 1 $eV$ and -1 $eV$ energy range. From the earlier available reports\cite{Winterlik1,Ramesh Kumar,Ming}, one saddle point is observed in Pd based compounds, at L point. But in the case of present compounds, we have two saddle points in both valence and conduction regions near the E$_F$. The flat bands associated with the van Hove singularity at the L-point result in a maximum density of states.

	We have also plotted the Fermi surface (FS) for the bands which are crossing E$_F$ and are given in Fig 4. We have one FS in both Ni$_2$NbAl and Ni$_2$NbGa compounds which is of electron nature due to one band crossing the E$_F$ at X point. In the case of Ni$_2$NbSn two bands cross E$_F$, the blue coloured band (indicated with dotted line) cross the E$_F$ from conduction band to valence band indicating the electron nature of the band and the remaining red coloured band (indicated with breaking line) cross the E$_F$ from conduction band to valence band at X point and again from valence to conduction band at W points indicating the mixed character of the band. In a similar way, we have two FS both having electron nature  in Ni$_2$VAl. In all the compounds we have observed parallel sheets in FS which indicate a nesting feature.

Now we briefly discuss the mechanical properties. The calculated elastic constants satisfy the Born mechanical stability criteria \cite{BORN} i.e. $C_{11}>0$, $C_{44}> 0$, $C_{11} > C_{12}$, and $C_{11} + 2C_{12}> 0$ as expected. In Table-II, we have given the single crystalline elastic constants along with calculated Debye temperature values. The polycrystalline elastic constants can be calculated from the single crystalline elastic constants the relations between single and polycrystalline elastic constants can be found elsewhere \cite{V.KANCHANA,VGAAM,VGAA,VGA}.
From the same table the calculated Debye temperature ($\theta_D$) values agree well with experiments. BCS theory predicts that T$_c$ should increase with increasing frequency of the lattice vibrations. For some Heusler compounds \cite{Klimczuk} T$_c$ decreases with the  increasing $\theta_D$. In the case of Ni$_2$NbX where X= Al, Ga, both T$_c$ and $\theta_D$ show the BCS theory behaviour. In the case of Ni$_2$NbSn it is showing opposite manner. In Pd based Huesler compounds Klimzuk et. al\cite{Klimczuk} found a decrease in T$_c$ with increasing $\theta_D$. In their study, T$_c$ of (Sc, Y, Lu)Pd$_2$Sn compounds increases with N(E$_F$) and found the opposite trend in case of APd$_2$M (A = Zr, Hf; M = Al, In) compounds. Finally they conclude that the change in T$_c$ is dependent on the system.

\subsection{Vibrational and superconducting properties}

The calculated phonon dispersion curves are shown in Fig.5 along with the total and partial phonon density of states. The primitive unit cells of the present compounds have one formula unit with four atoms which gives 12 phonon branches including three acoustic and nine optical branches. The absence of imaginary phonon frequencies indicates dynamical stability at the ambient condition. The higher frequency optical phonon modes are separated from others in Ni$_2$NbAl, Ni$_2$NbGa and Ni$_2$VAl compounds. But the same is not found in the case of Ni$_2$NbSn. This separation is due to the mass difference between different kind of atoms in the unit cell. In Ni$_2$NbAl and Ni$_2$VAl these higher frequency optical modes are due to the lighter Al atom. In the remaining compounds, it is due to Nb atom. Again in Ni$_2$NbAl and Ni$_2$VAl, acoustic and lower frequency optical modes are due to the vibrations of Ni atoms. In all the compounds at the zone centre, we have three optical phonon modes and in that one $T_{2g}$ mode is Raman active and two $T_{1u}$ modes are infrared active. Frequency of these $T_{2g}$ mode in Ni$_2$NbX (X = Al, Ga, Sn) and Ni$_2$VAl is 139.8, 148.9, 139.7 and 138.0 $cm^{-1}$ respectively at $\Gamma$ point. In all the compounds we have doubly degenerate acoustic and optical modes from $\Gamma$-X and triple degeneracy of same modes is observed along X-$\Gamma$ direction. This degeneracy is due to the symmetry of the crystal in cubic phase. In all the compounds longitudinal acoustic mode (LA) is interacting with $T_{2g}$ optical modes. We have observed degenerate transverse acoustic (TA) modes in $\Gamma$ - X and $\Gamma$ - L directions in all the compounds. In other directions these TA modes become non degenerate  and split into TA1 (high frequency) and TA2 (low frequency) modes. There is an anomaly (dip) in TA2 mode from X-$\Gamma$ direction in all the compounds except Ni$_2$NbSn. In the case of Ni$_2$NbSn, it is observed at X high symmetry point. As discussed below these dips are related to the Fermi surface and reflect the electron phonon coupling.

In metals such dips can arise from Fermi surface nesting, i.e. Kohn anomalies in the phonon spectrum. This anomaly is observed in Ni$_2$MnGa \cite{Bungaro,Zayak,Ayuela}, Ni$_2$MnIn \cite{Agduk} and Ni$_2$MnX (X= Sn,Sb) \cite{Agduk1}. From the phonon dispersion of the present compounds we observed dip (softening) in the lower frequency acoustic mode (TA2) along X-$\Gamma$ direction in both Ni$_2$NbAl, Ni$_2$NbGa and Ni$_2$VAl compounds and near the X-point in Ni$_2$NbSn. From the FS, we observed flat portions of X- point Fermi surfaces in all the investigated compounds to be nested with the similar portion on the other side. FS with nesting vector direction is given in Fig.6(a). From this we observed that the nesting vector is of length $\sim$0.61 of X-X distance connecting the flat surfaces of FS. In the case of Ni$_2$NbGa and Ni$_2$VAl it is at $\sim$0.64 and $\sim$0.67 of X-X distance. The nesting vector is around 0.7 $\frac{2\pi}{a}$ along the $\Gamma$-X direction in Ni$_2$NbAl. In case of  Ni$_2$NbGa and Ni$_2$VAl it is observed at 0.68 $\frac{2\pi}{a}$, 0.62 $\frac{2\pi}{a}$ respectively in the same direction as Ni$_2$NbAl. The same nature is observed in Ni$_2$NbSn at X high symmetry point. To know the exact 'q' vector where the anomaly is observed, we have plotted the acoustic phonon modes form  $\Gamma$ to X direction for Ni$_2$NbAl, Ni$_2$NbGa and Ni$_2$VAl compounds as shown in Fig. 6(b,c,d), where we observed a strong Kohn anomaly at 
$\xi$ = 0.7, 0.68 and 0.62 in Ni$_2$NbAl, Ni$_2$NbGa and Ni$_2$VAl respectively. The associated phonon branches have mainly Ni character. In the case of Ni$_2$MnGa, Ni$_2$MnSn and Ni$_2$MnSb compounds also the same type of atoms are involved in the branches showing Kohn anomalies.

The electron phonon coupling constants ($\lambda_{ep}$) were extracted from the Eliashberg function ($\alpha^2$F($\omega$)) which can be used to determine the T$_c$ of a conventional phonon mediated superconductor. The calculated $\alpha^2$F($\omega$) are plotted in Fig. 7 for all the investigated compounds. From this figure the lower energy phonon modes, which are mainly due to Ni atom at that particular frequency are more involved in the process of scattering the electrons in all the compounds. The T$_c$ of the present compounds is calculated by using Allen-Dynes \cite{Allen} formula which is given in equation below,

\begin{equation}
T{_c}=\frac{\omega_{ln}}{1.2}exp(-\frac{1.04(1+\lambda_{ep})}{\lambda_{ep}-\mu{^*}(1+0.62\lambda_{ep})})
\end{equation}
where $\omega_{ln}$ is logarithmically averaged phonon frequency, $\lambda_{ep}$ is electron phonon coupling constant and $\mu^*$ is Coulomb pseudopotential which is a parameter that normally takes a value of 0.1-0.15.

The electron phonon coupling constant ($\lambda_{ep}$) is usually extracted from the Eliashberg function ($\alpha^2F(\omega)$) which can be used to determine the $T_c$ of a conventional phonon mediated superconductor. $\alpha^2F(\omega)$ is 

\begin{equation}
\alpha^2F(\omega)=\frac{1}{2\pi N(\epsilon_f)}\sum_{qj} \frac{\nu_{qj}}{\hbar\omega_{qj}}\delta(\omega-\omega_{qj})
\end{equation}

This function is related to the phonon DOS ($F(\omega) = \sum_{qj} \delta(\omega-\omega_{qj})$) and differs from the phonon DOS by having a weight factor $1/2\pi N(\epsilon_f)$ inside the summation. In the above formula $N(\epsilon_f)$ is the electronic density of states at the E$_F$ and $\nu_{qj}$ is the phonon line width which can be represented as below.  
\begin{equation}
\nu_{qj}= 2\pi\omega_{qj}\sum_{knm}|g_{(k+q)m,kn}^{qj}|^2\delta(\varepsilon_{kn}-\varepsilon_F)\delta(\varepsilon_{(k+q)m}-\varepsilon_F)
\end{equation}

where Dirac delta function express the energy conservation conditions and $g$ is the electron phonon matrix element. The electron phonon coupling constant ($\lambda_{ep}$) can be expressed in terms of $\alpha^2F(\omega)$ as shown below.

\begin{equation}
\lambda_{ep}=2 \int\frac{d\omega}{\omega}\alpha^2F(\omega)
=\int\lambda(\omega) d\omega
\end{equation}

where $\lambda(\omega)= \frac{2\alpha^2F(\omega)}{\omega}$

The calculated T$_c$ values for all the investigated compounds are given in Table-III for $\mu^*$ values of 0.13 and 0.15. The calculated values agree well with the experiments with $\mu^*$ = 0.13 for Ni$_2$NbX (X= Al, Ga, Sn) compounds. Among the investigated compounds, Ni$_2$VAl is found to have relatively high T$_c$ of 3.84 $K$ with high $\lambda_{ep}$ value 0.68 with $\mu^*$ = 0.13. As we discussed Ni$_2$VAl has high electronic DOS and high Sommerfield coefficient $\gamma$ at E$_F$ compared to other compounds indicating the compound to be a superconductor with high T$_c$ value than the other compounds. From the calculated T$_c$ and $\lambda_{ep}$ values, the superconducting nature of the Ni$_2$VAl compound is evident. These calculated values are in the range of the present Ni$_2$NbX (X= Al, Ga, Sn) and other Ni based superconducting Heusler compound ZrNi$_2$Ga\cite{Ming}. 
 
\subsection{Under compression}

As mentioned, pressure is a clean tuning parameter for exploring the relationships between superconductivity and other physical properties. 
This motivated us to proceed further to see the effect of pressure on the above mentioned properties for the compounds studied. For all the compounds we compressed the volume up to -15\% of the initial volume. The pressure effect on the band structure and FS of Ni$_2$NbAl compound is already reported in our previous work \cite{PVSR}, where we find an extra band to cross the E$_F$ at the compression of V/V$_0$=0.93 (pressure of 17 $GPa$) at X point and due to this we have an extra FS at the same point having the electron character. In the case of remaining two compounds, we have not observed any change in band structure and FS topology under compression as mentioned above. 

The electronic DOS under compression and the values are shown in Fig. 8(a). From the plot, the total electronic DOS linearly decreases with pressure in Ni$_2$NbGa, Ni$_2$NbSn and Ni$_2$VAl compounds but in the case of Ni$_2$NbAl it is non-linear at V/V$_0$ = 0.93 and is clearly represented in the inset.

We have also calculated the single crystalline elastic constants for all the compounds under compression to check the effect of pressure on the mechanical stability in the present compounds and are plotted in Fig 8(b). From this we can observe that all the compounds are satisfying the Born's \cite{BORN} stability criteria under compression indicating the mechanically stable nature of the present compounds under the compression range we have studied. We also observed that the values of three independent elastic constants increases under compression in all the compounds as usual. It is also observed that $C_{11}$, $C_{12}$ are more sensitive to pressure while $C_{44}$ is quite insensitive to pressure for the same compounds, where $C_{44}$ is related to transverse distortion which is almost flat indicating the effect of pressure on this to be weaker.

We calculated the phonon dispersion under compression. We find hardening of frequencies for all the modes except in the lowest frequency acoustic mode which softens under compression in all the compounds. In Fig 9, we have given only the lower frequency acoustic mode under compression for all the compounds to show the softening nature in that particular mode. From this figure, we have observed that the softening becomes more pronounced under compression at the same 'q' vector where the Kohn anomaly is found in all compounds. The softening in phonon frequency, corresponding to the Kohn anomaly, under pressure in the Nb compounds implies a pressure dependent structural phase transition. Near the transition pressure this is presumably of charge density wave character. In Ni$_2$NbAl it is observed to be imaginary at compression V/V$_0$= 0.92 and 0.85. In this compound, we find an extra Fermi surface at V/V$_0$ = 0.93. The change in the FS topology could be reason for the imaginary mode at V/V$_0$ = 0.92 compression. And also the nesting feature under compression becomes more prominent due to increase in the size of the FS under compression. In the remaining compounds, no extra FS is observed but the size of the FS increases under compression which might lead to an increase in the effect of nesting under compression. In Ni$_2$MnSb\cite{Agduk1} authors observed the imaginary frequency in TA2 mode at ambient condition which is due to the presence of Khon anomaly in this system. The imaginary frequencies under compression in TA2 mode in all the compounds may be due to the presence of same Kohn anomaly  under compression. This acoustic mode softening is also observed in other Heusler compounds. In HfPd$_2$Al\cite{Wiendlocha} it is observed at the pressure of 7.5 GPa, Pd$_2$ZrAl\cite{Winterlik1} at ambient condition. In the case of YPd$_2$Sn\cite{Tutuncu}, authors found anomaly in the transverse acoustic mode and reported that to be the reason for the  increase in electron-phonon coupling parameter of that phonon branch. The anomaly in the transverse acoustic mode is also observed in non Heusler compounds\cite{Isaev,Isaev1,Bagci,Bagci1,Tutuncu1}, where the authors reported that these phonon anomalies play an important role in understanding superconductivity in those compounds. As discussed previously, we have observed the anomalies in the phonon frequencies at X-$\Gamma$ point in Ni$_2$NbAl, Ni$_2$NbGa and Ni$_2$VAl compounds and at X point in Ni$_2$NbSn. We already know that the lower frequency phonon modes would contribute more to the electron phonon coupling, which further has an impact on the T$_c$ of that material. This indicate that the softening may lead to change in the $\lambda_{ep}$ and T$_c$ in these compounds.

	We have calculated the electron-phonon coupling constant and superconducting transition temperature for all the compounds under compression and are shown in Fig 10. From these plots, all the compounds show a non-monotonic variation in T$_c$ and the electron-phonon coupling constant under compression are behaving in the opposite manner to $\omega_{ln}$. In prior work\cite{P B Allen, H Wuhl} it was reported that the softening of the phonon DOS leads to the increase in the T$_c$ of that material. 
In Ni$_2$NbAl, the T$_c$ plot under compression in our previous work \cite{PVSR} is different from this work. In our previous work we used ultrasoft pseudopotentials and also we did not find any softening nature in the phonon dispersion curve as well as phonon density of states. In the present work we have used norm-conserving pseudopotentials. These are more difficult to converge and require more computational resources but are more reliable.

\section{Conclusions}

 We have studied the ground state, electronic structure, mechanical, vibrational and superconducting properties of Ni$_2$NbX (X = Al, Ga and Sn) and Ni$_2$VAl Heusler compounds using the density functional theory within a generalised gradient approximation at ambient and under compression. Our calculated ground state properties agree well with experimental and other theoretical results. From these calculations, at the E$_F$ the contribution in DOS is mainly from Ni $d_{eg}$ states with the admixture of Nb/V $d_{t_{2g}}$ states in all the compounds. The Fermi surface resulting from the bands crossing the E$_F$ are also shown and observed that one electron pocket exist in both Ni$_2$NbAl and Ni$_2$NbGa compounds and two FS of electron nature are found in the case of Ni$_2$NbSn and Ni$_2$VAl compounds. Nesting features in FS are evident in all the compounds.  The calculated single crystalline elastic constants satisfy the Born stability criteria indicating the mechanical stability of the investigated compounds both at ambient and under compression. The absence of the imaginary frequencies in phonon dispersion calculations indicate the dynamical stability of the present compounds. Kohn anomaly is observed in TA2 mode in all the compounds which might be due to the nesting in the FS. The computed T$_c$ using Allen-Dynes formula agree well with the experimental results. Importantly, Ni$_2$VAl is predicted to be a superconductor which have T$_c$ value in the range of Heusler compounds. From the calculated Eliashberg function ($\alpha^2$F($\omega$)) we have observed that Ni atom contribution is more towards T$_c$ in all the compounds. Under compression we have observed the change in the band structure of Ni$_2$NbAl compound where we find an extra band to cross the E$_F$. In addition we also observed a pronounced softening  in the Kohn anomalies existing in TA2 mode under compression. This phonon softening again lead to softening in the phonon density of states. A non-monotonic variation in the T$_c$ under compression is notified which is due to softening in the lower frequency acoustic mode and phonon DOS. Among the compounds studied, we observe that the variation in T$_c$ under pressure is minimal in Ni$_2$VAl. It will be of interest to experimentally search for superconductivity in Ni$_2$VAl and study its pressure dependence.

\section{Acknowledgement}

	The authors P.V.S.R and V.K would like to thank Department of Science and Technology (DST) for the financial support through SR/FTP/PS-027/2011. P.V.S.R would also like to acknowledge IIT-Hyderabad for providing computational facility. G.V would like to acknowledge CMSD-University of Hyderabad for providing computational facility. V.K and G.V would like to acknowledge Axel Svane and N. E. Christensen for fruitful discussions during the beginning of this project.


\newpage

\begin{table}
\caption{\label{arttype}Ground state properties of Ni$_2$NbX (X = Al, Ga, Sn) and Ni$_2$VAl at ambient pressure combined with experimental reports. where $a_{exp}$ is experimental lattice parameter, $a_{th}$ is theoretical lattice parameter, $B$ is bulk modulus, $\gamma_{exp}$ is experimental Sommerfield coefficient,  $\gamma_{th}$ theoretical Sommerfield coefficient.}

\footnotesize\rm
\begin{tabular*}{\textwidth}{@{}l*{15}{@{\extracolsep{0pt plus12pt}}l}}
\br

\begin{tabular}{cccccccc} 
 Parameters &Ni$_2$NbAl &Ni$_2$NbGa &Ni$_2$NbSn &Ni$_2$VAl\\
\mr
$a_{exp}$ (\AA) &5.969\cite{S Waki},5.9755\cite{Rocha} &5.956\cite{S Waki} &6.157\cite{S Waki}, 6.160\cite{Boff}, 6.179\cite{JH Wernick} &5.8031\cite{Rocha}, 6.33\cite{Villars}\\

 $a_{th}$ (\AA) 	&5.996, 6.00\cite{Lin} &5.991 &6.202, &5.800, 5.78\cite{Lin} \\   	            

 $B$ ($GPa$) 		&181 &182 &170&181\\

 $\gamma_{exp}$ ($mJ/mol K^{2}$) &8.0\cite{S Waki}, 11.03\cite{Rocha} &6.5\cite{S Waki} &4.0\cite{S Waki}, 5.15\cite{Boff} &14.18\cite{Rocha}\\

 $\gamma_{th}$ ($mJ/mol K^{2}$) &5.36 &5.19 &5.52 &8.27\\

\br
\end{tabular}
\end{tabular*}
\end{table}

\begin{table*}
\caption{\label{arttype}Calculated single crystalline elastic constants and Debye temperature at ambient condition for Ni$_2$NbX (X = Al, Ga, Sn) and Ni$_2$VAl.}
\footnotesize\rm
\begin{tabular*}{\textwidth}{@{}l*{15}{@{\extracolsep{0pt plus12pt}}l}}
\br
\begin{tabular}{ccccccccc}
Parameters &Ni$_2$NbAl &Ni$_2$NbGa &Ni$_2$NbSn &Ni$_2$VAl\\
\mr
$C_{11}$ (GPa)		&212	&194	&188	&200\\	
$C_{12}$ (GPa)		&167	&176	&162	&171\\
$C_{44}$ (GPa)		&98	&95	&72	&109\\
$\Theta_D$ (K)		&306	&241	&219	&319\\
$\Theta_D$(experimental)(K) &280\cite{S Waki}, 300\cite{Rocha} &240\cite{S Waki} &206\cite{S Waki}	&358\cite{Rocha}\\	
\br
\end{tabular}
\end{tabular*}
\end{table*}

\begin{table}
\caption{\label{arttype}Calculated T$_c$ and $\lambda_{ep}$ values with experimental reports for Ni$_2$NbX (X = Al, Ga, Sn) and Ni$_2$VAl.}
\footnotesize\rm
\begin{tabular*}{\textwidth}{@{}l*{15}{@{\extracolsep{0pt plus12pt}}l}}
\br
\begin{tabular}{cccccccc}

Parameters &Ni$_2$NbAl &Ni$_2$NbGa &Ni$_2$NbSn	&Ni$_2$VAl\\
\mr				
			       	      	   
$T_c$ (experimental) &2.15\cite{S Waki} &1.54\cite{S Waki} &2.90\cite{S Waki}, 3.4\cite{Boff}, 3.4\cite{JH Wernick}	&--\\
$T_c$ (this work with $\mu^*$ = 0.13)  & 1.92 &1.18 &3.21 &3.84\\
$T_c$ (this work with $\mu^*$ = 0.15)  & 1.40 &0.79 &2.60 &3.09\\
$\lambda$ (experiment) &0.52\cite{S Waki}, 0.514\cite{Rocha} &0.50\cite{S Waki} &0.61\cite{S Waki}	&--\\
$\lambda$ (this work) &0.56 &0.50 &0.68	&0.68\\
\br
\end{tabular}
\end{tabular*}
\end{table}

\begin{figure*}
\begin{center}
\includegraphics[width=80mm,height=80mm]{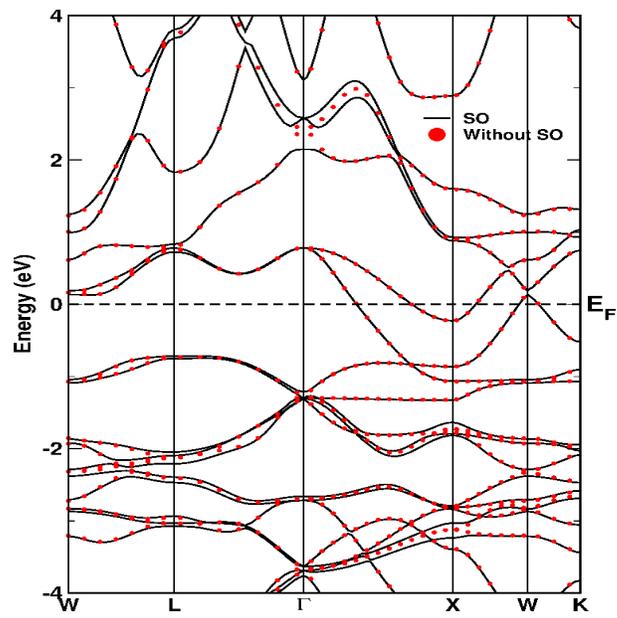}
\caption{(colour online)Band structure for Ni$_2$NbSn with and without inclusion of spin-orbit effect at the theoretical equilibrium volume.}
\end{center}
\end{figure*}

\begin{figure*}
\begin{center}
\subfigure[]{\includegraphics[width=60mm,height=60mm]{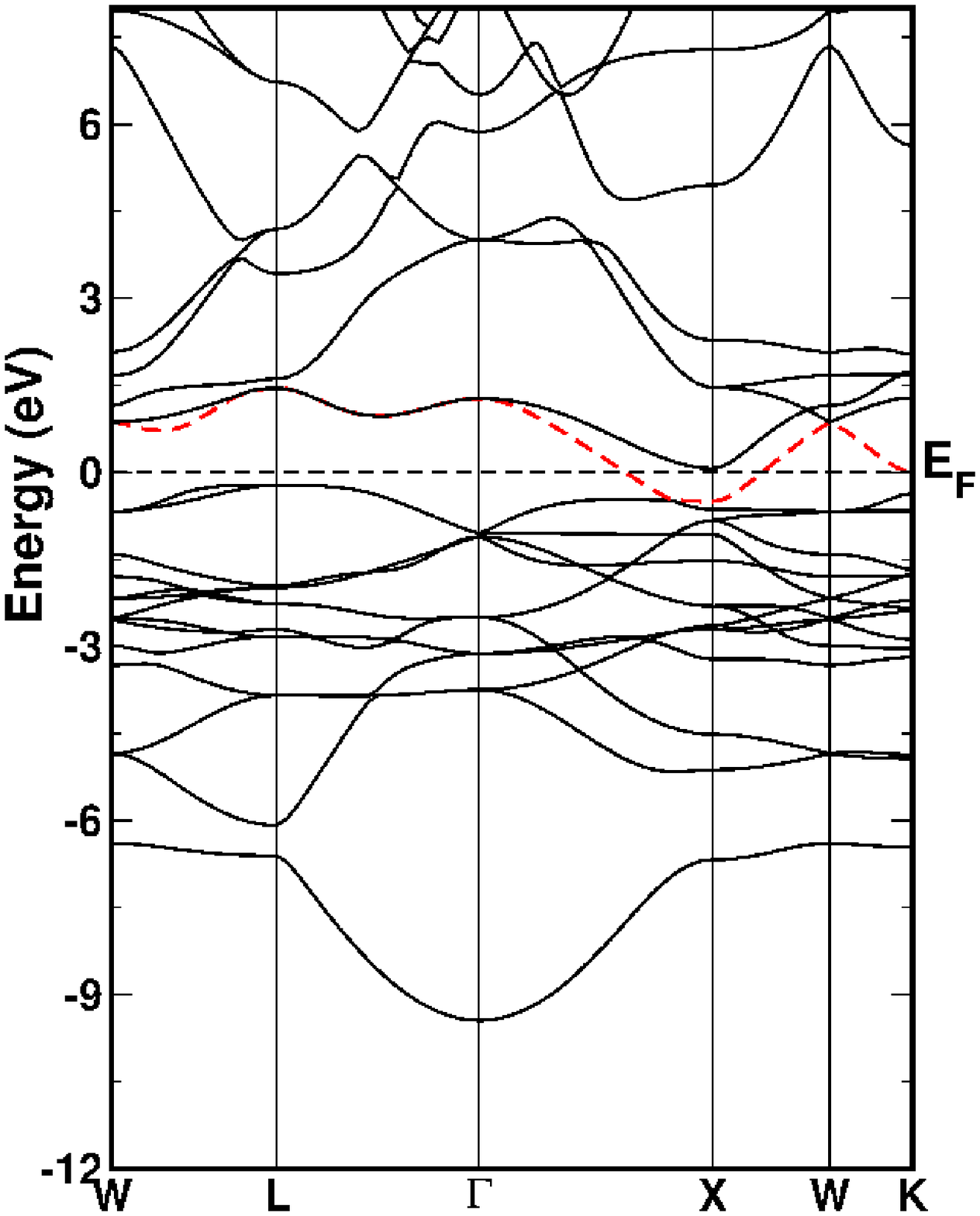}}
\subfigure[]{\includegraphics[width=60mm,height=60mm]{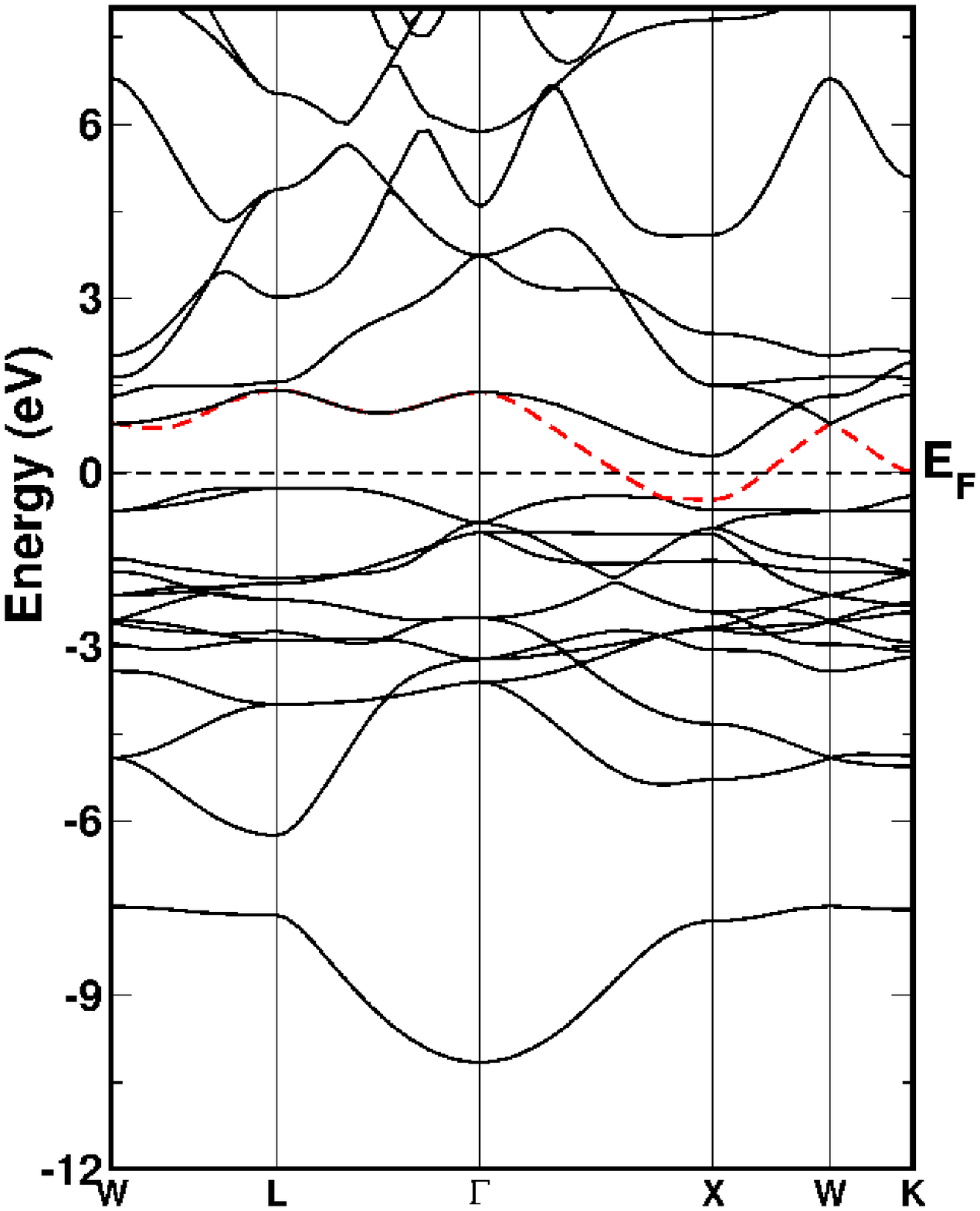}}
\subfigure[]{\includegraphics[width=60mm,height=60mm]{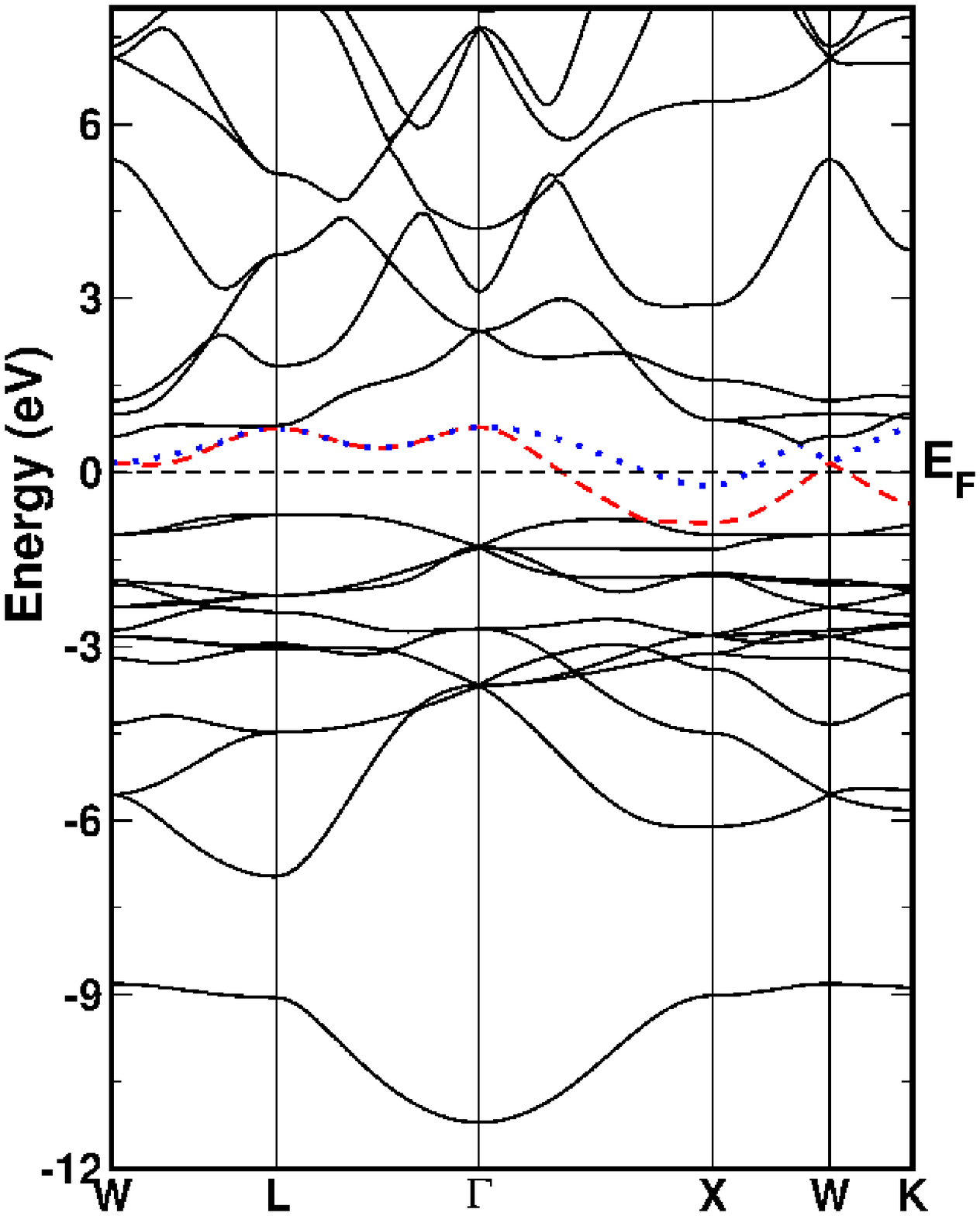}}
\subfigure[]{\includegraphics[width=60mm,height=60mm]{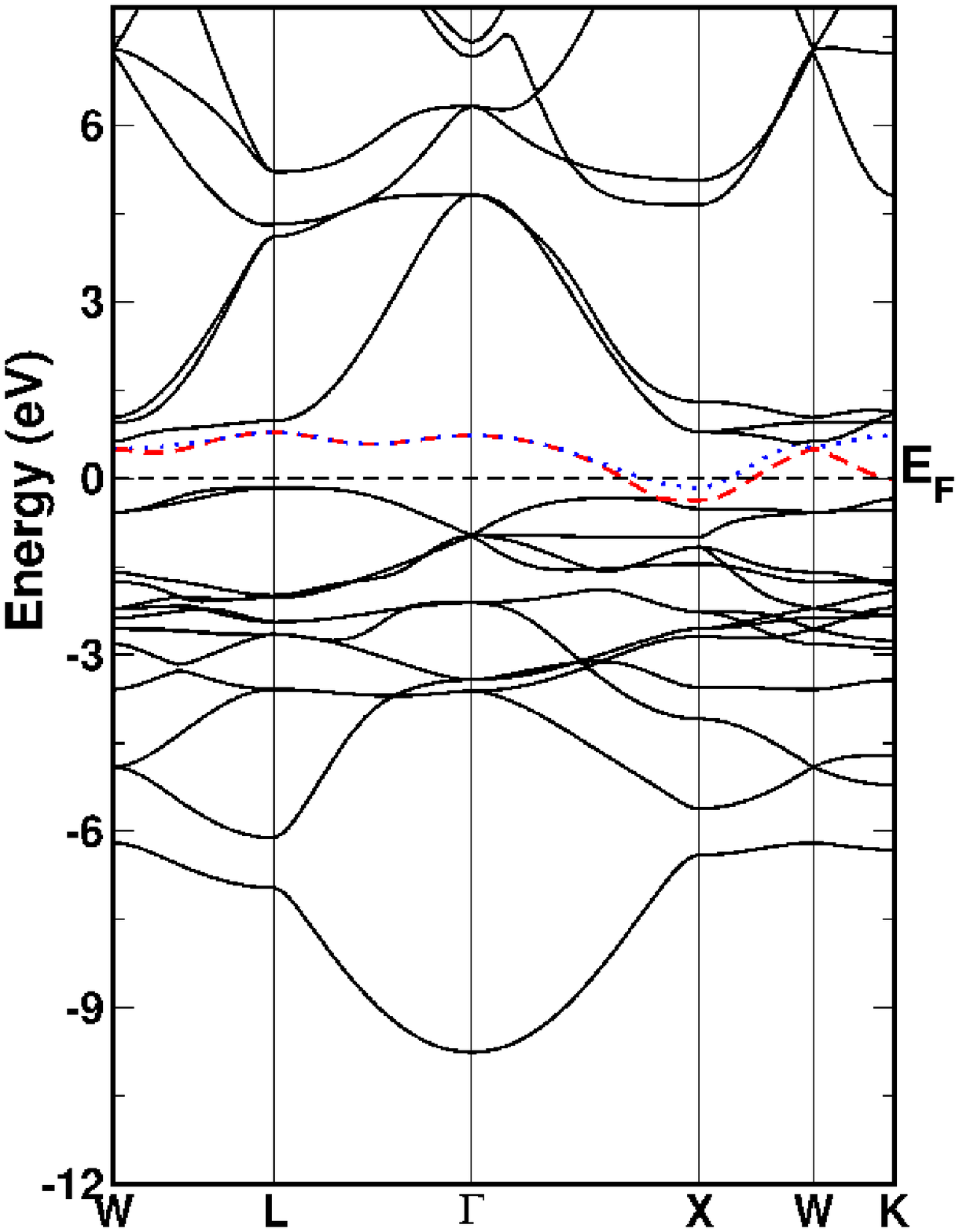}}
\caption{(colour online)Band structure for (a) Ni$_2$NbAl, (b) Ni$_2$NbGa, (c) Ni$_2$NbSn and (d) Ni2VAl .}
\end{center}
\end{figure*}

\begin{figure*}
\begin{center}
\subfigure[]{\includegraphics[width=70mm,height=70mm]{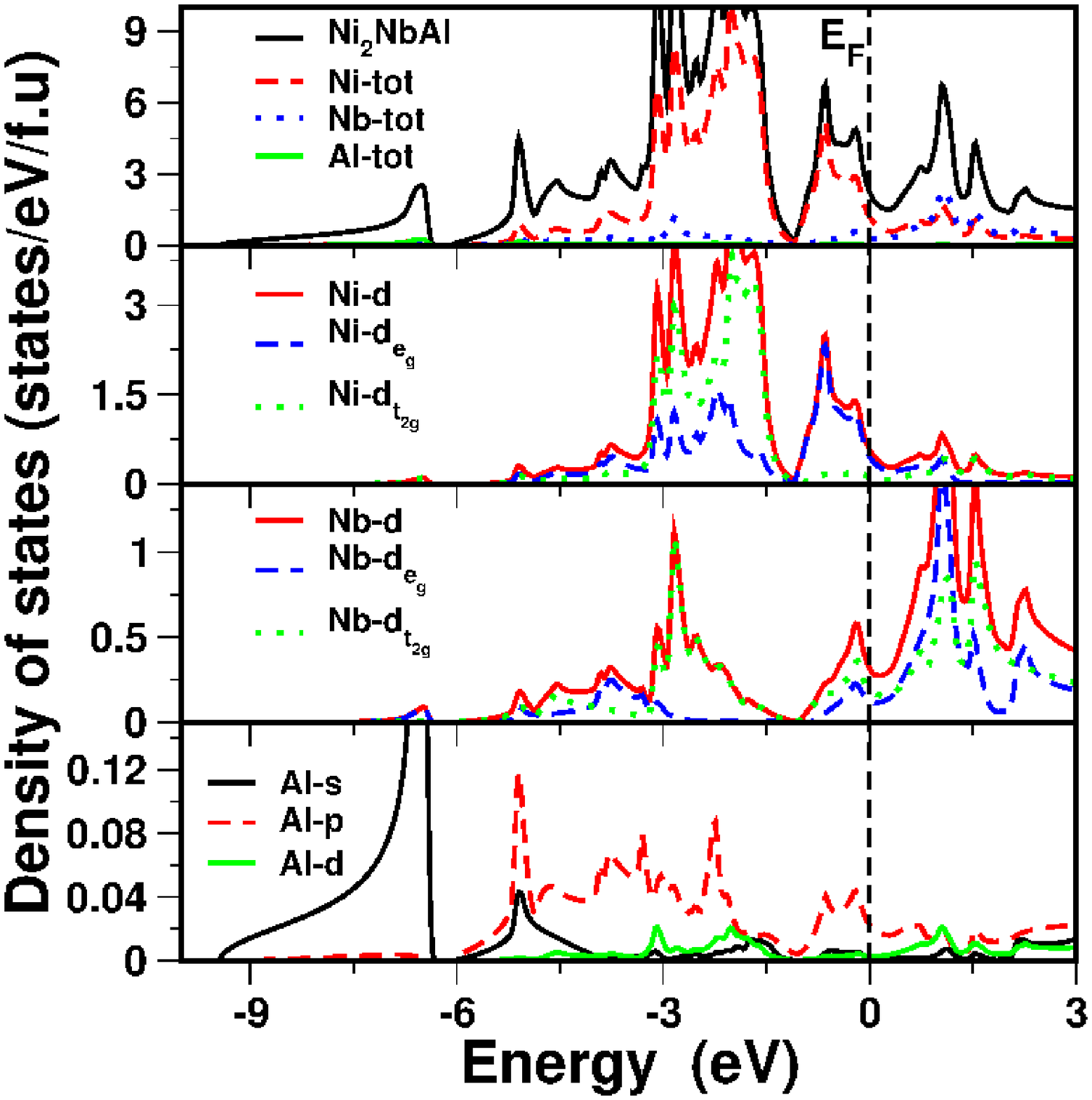}}
\subfigure[]{\includegraphics[width=70mm,height=70mm]{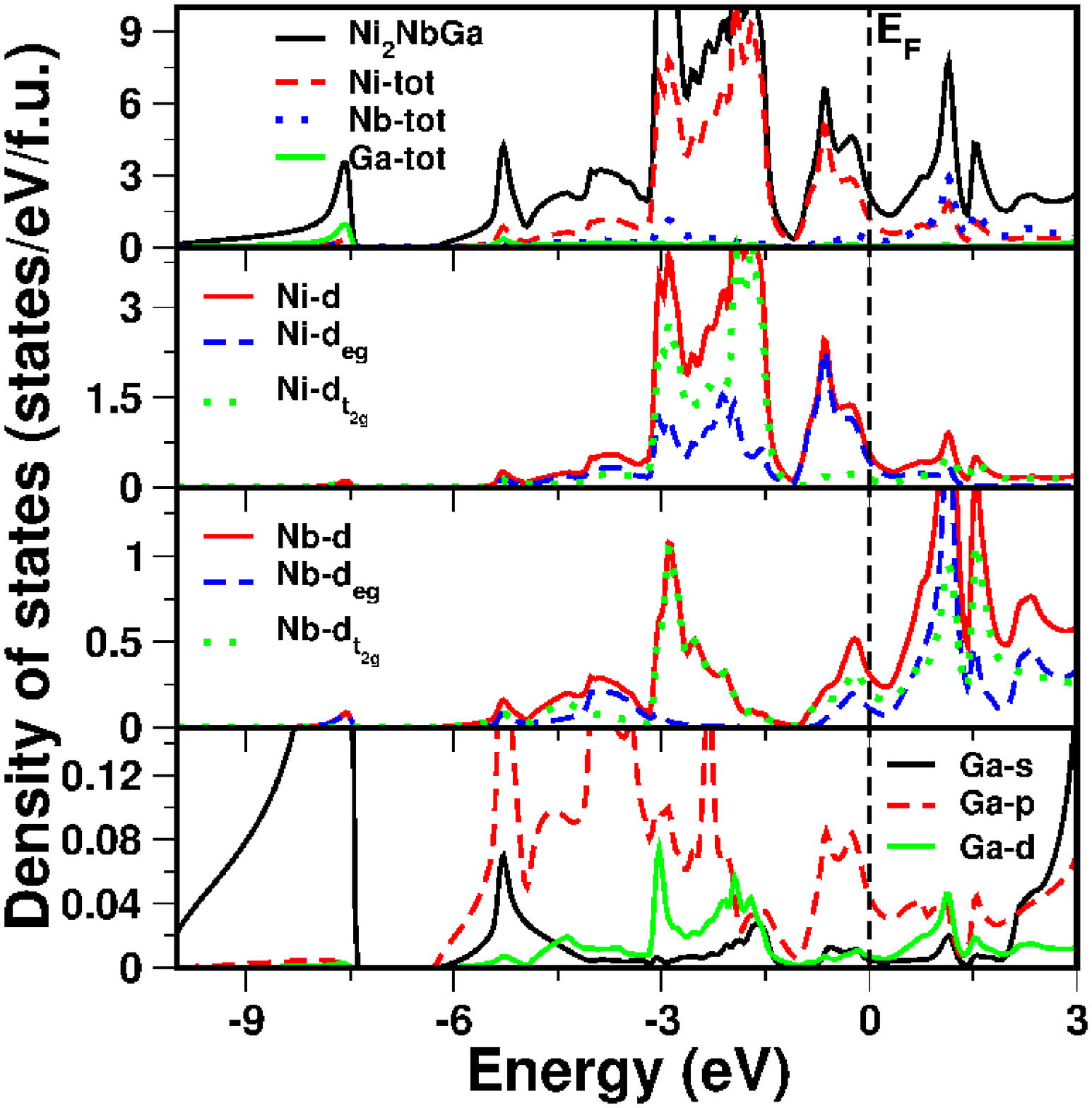}}
\subfigure[]{\includegraphics[width=70mm,height=70mm]{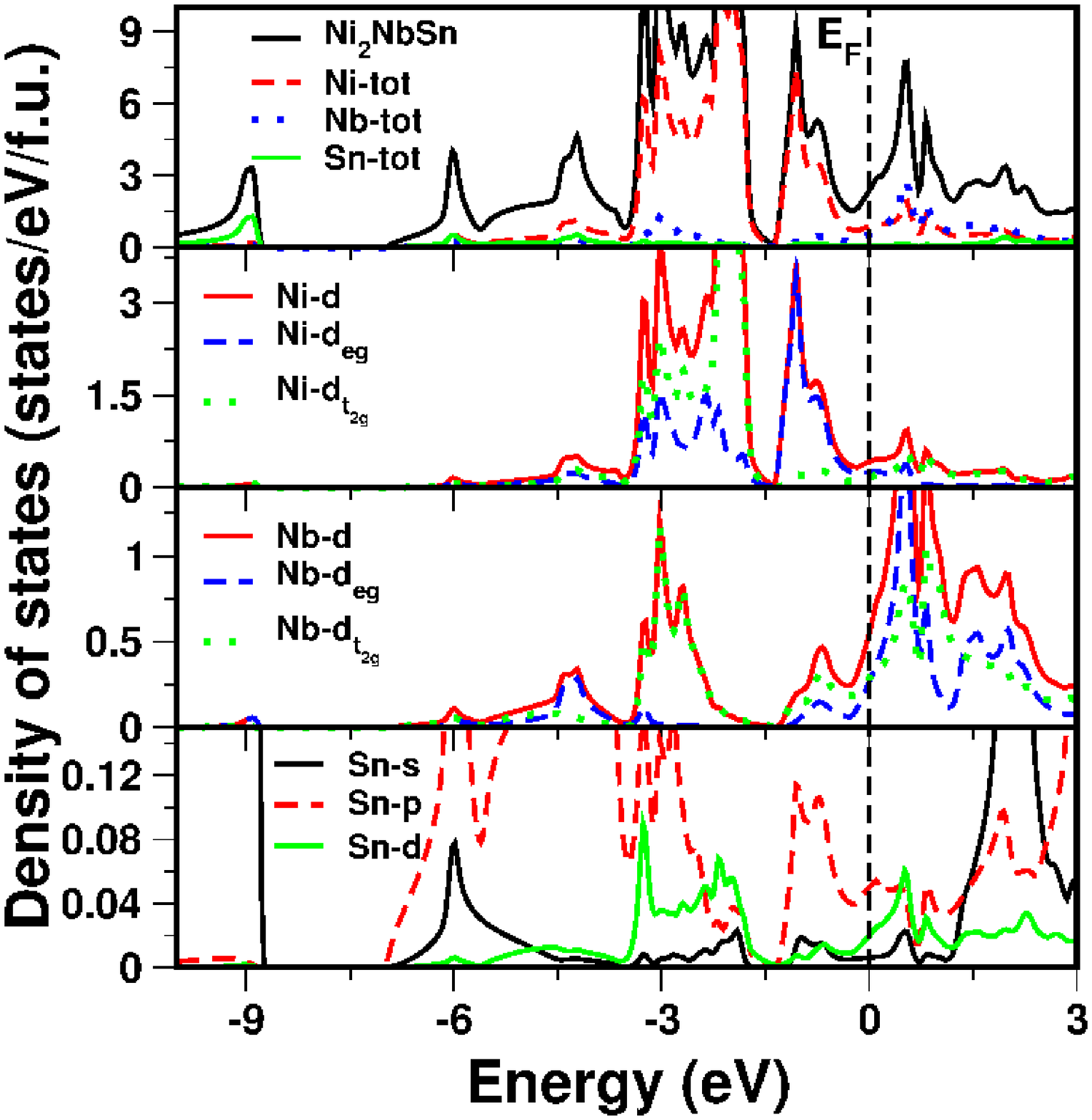}}
\subfigure[]{\includegraphics[width=70mm,height=70mm]{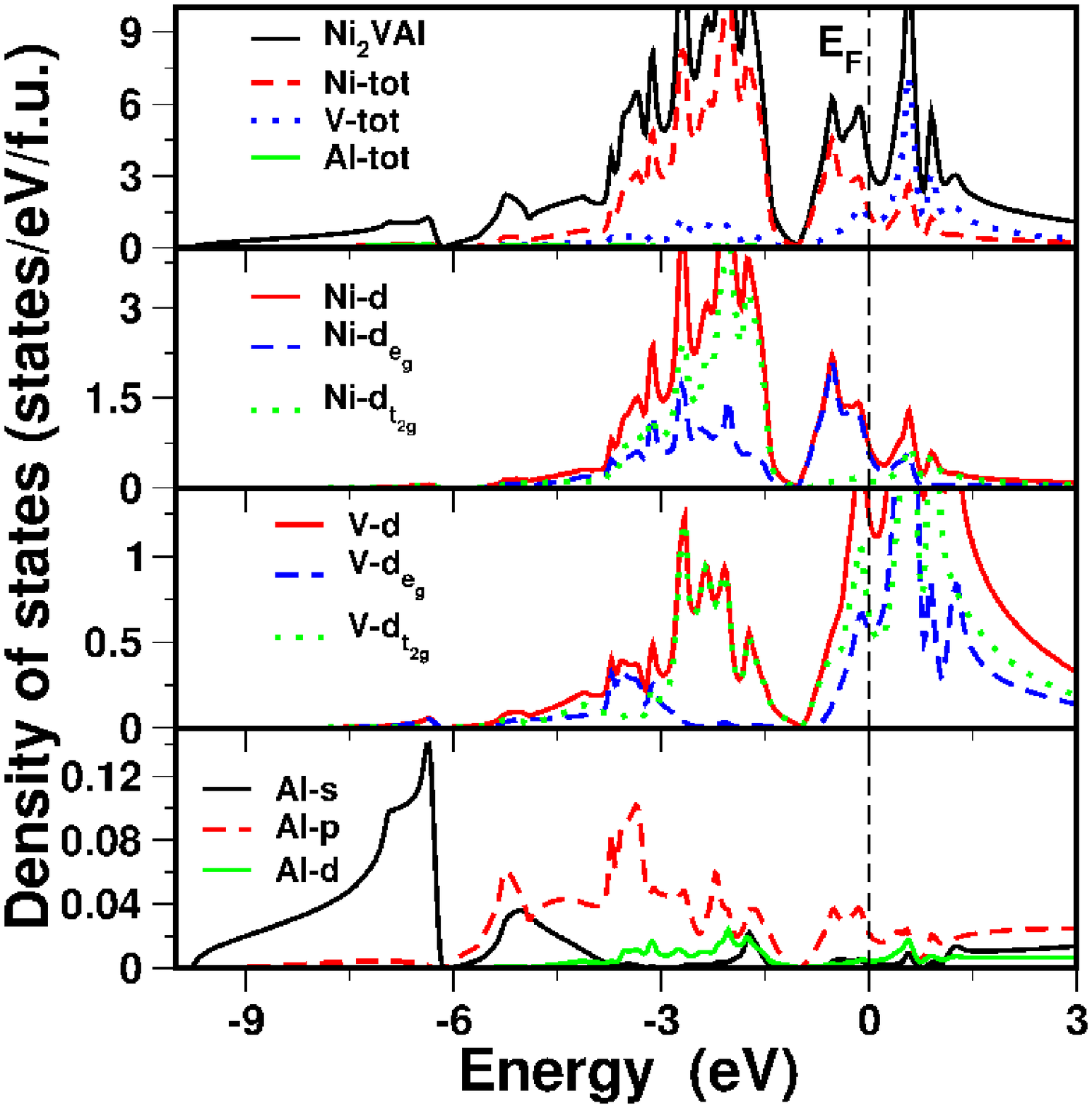}}
\caption{(colour online)Density of states for (a) Ni$_2$NbAl, (b) Ni$_2$NbGa, (c) Ni$_2$NbSn and (d) Ni2VAl.}
\end{center}
\end{figure*}

\begin{figure*}
\begin{center}
\subfigure[]{\includegraphics[width=35mm,height=35mm]{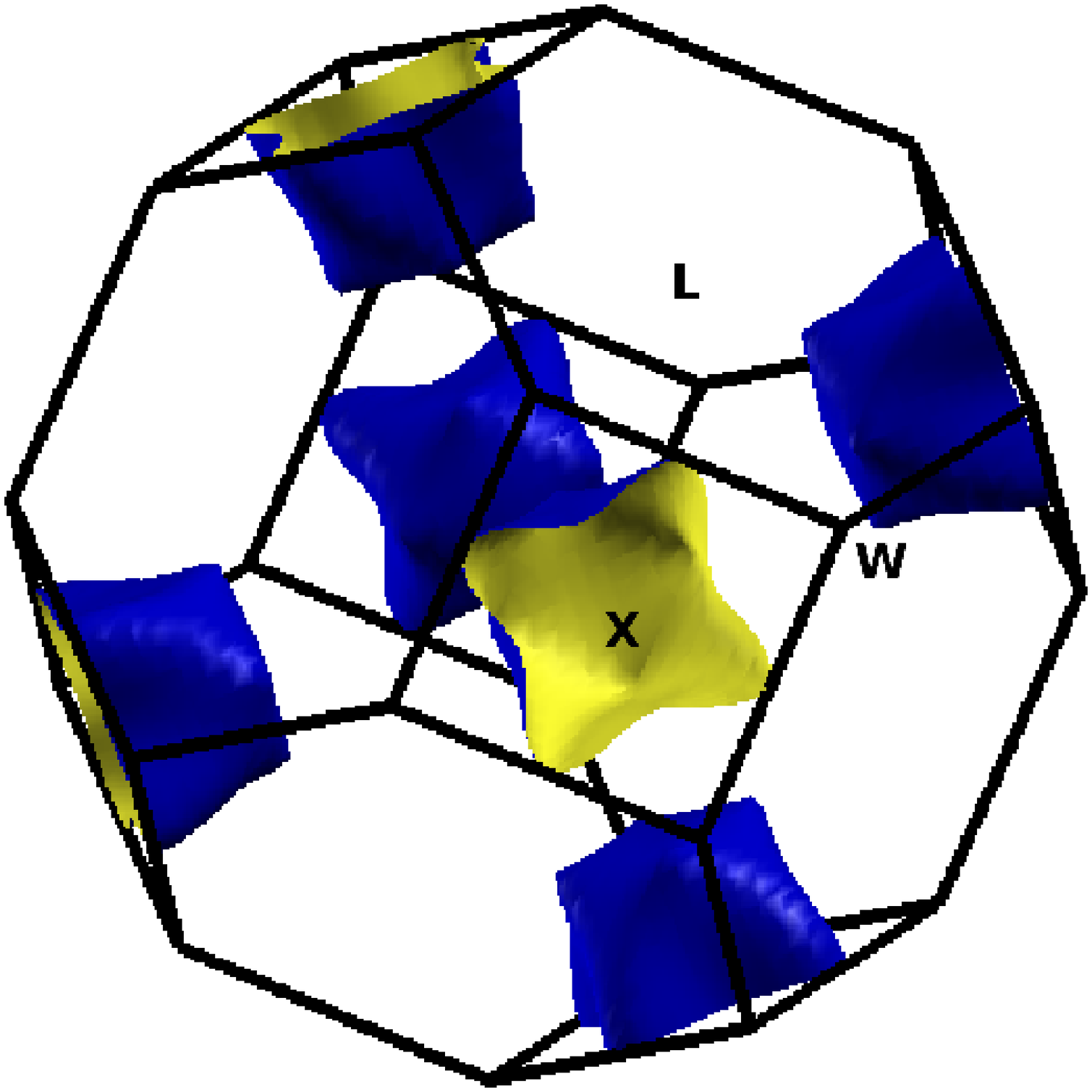}}
\subfigure[]{\includegraphics[width=35mm,height=35mm]{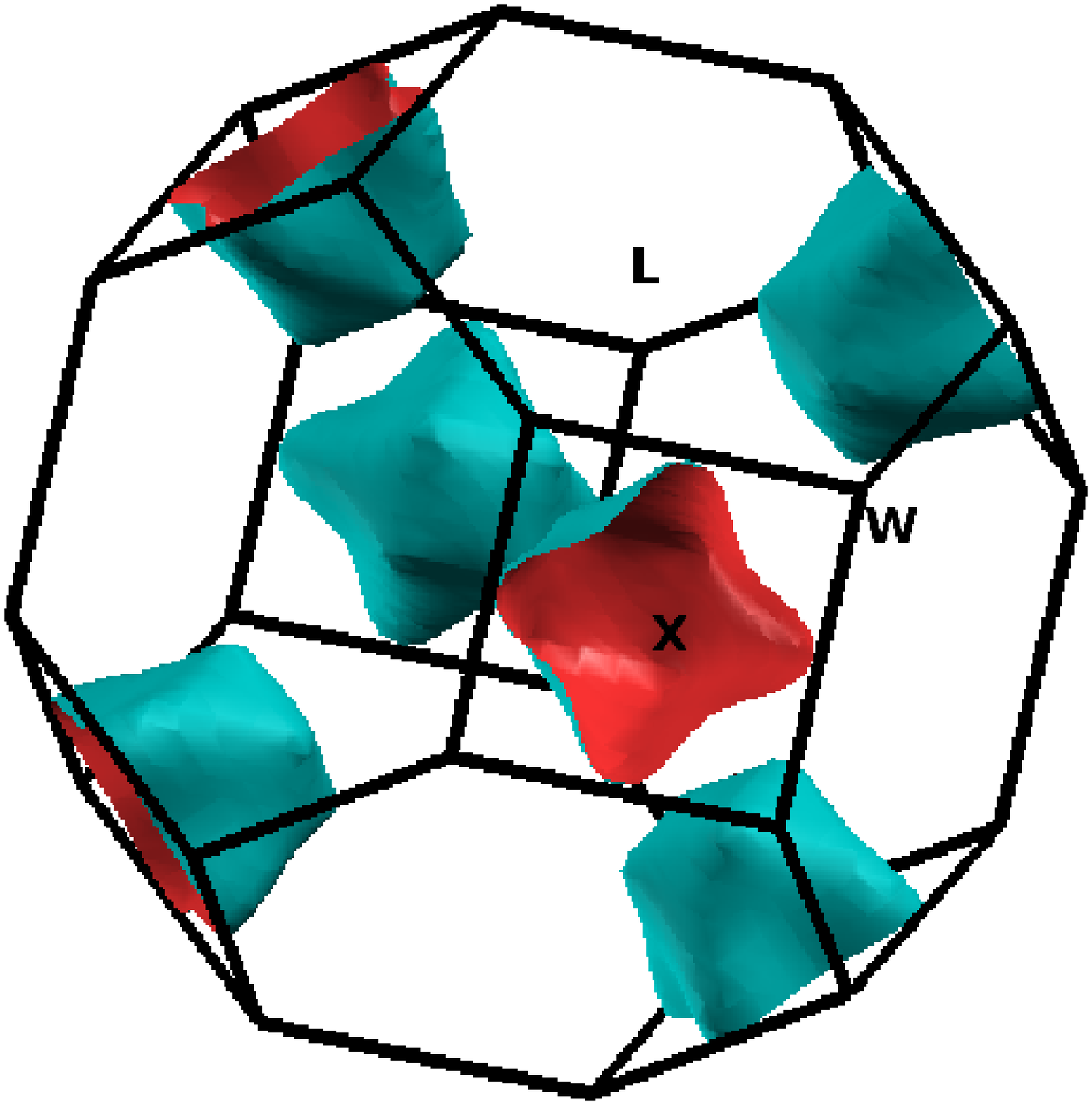}}
\subfigure[]{\includegraphics[width=35mm,height=35mm]{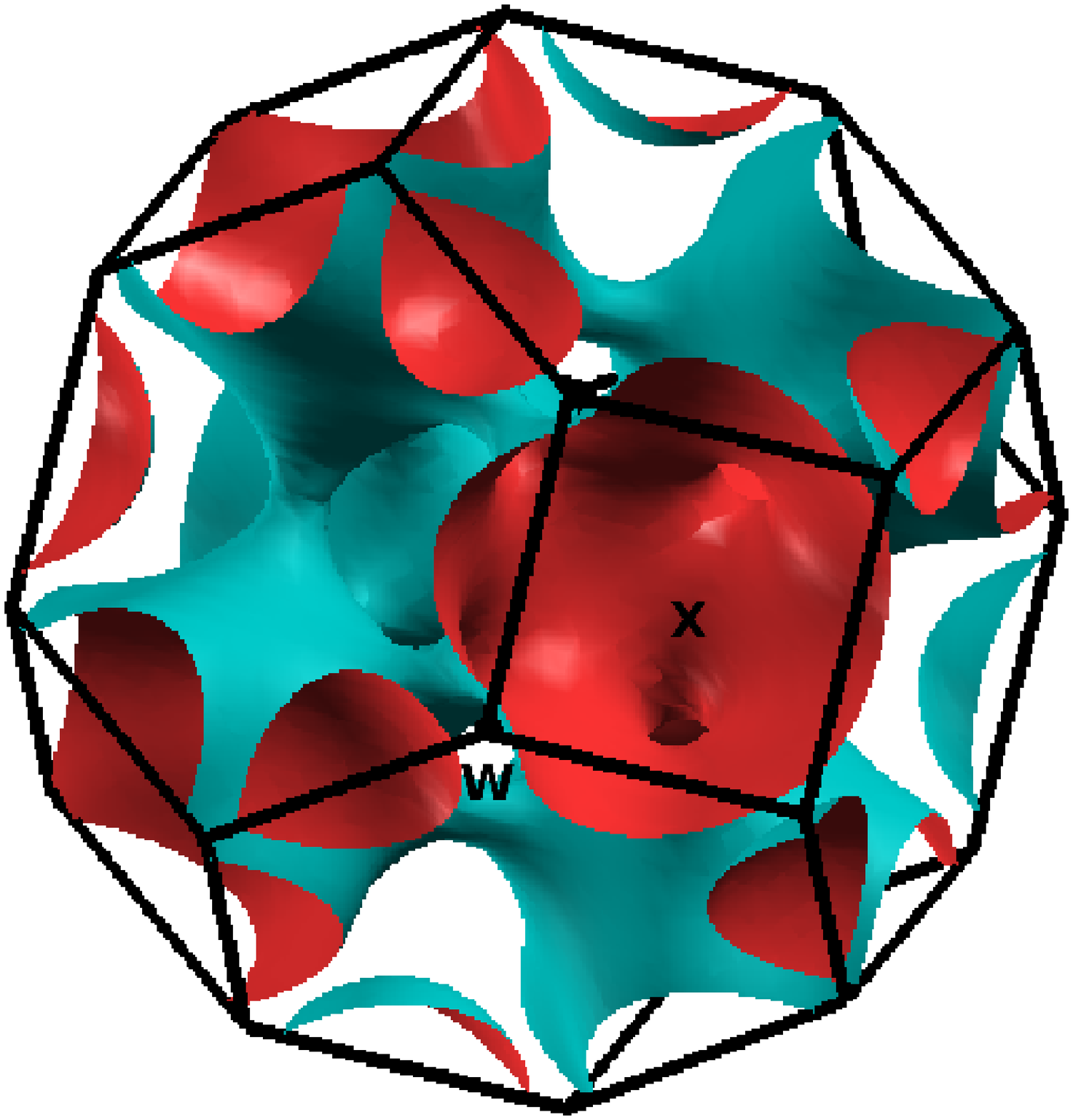}}
\subfigure[]{\includegraphics[width=35mm,height=35mm]{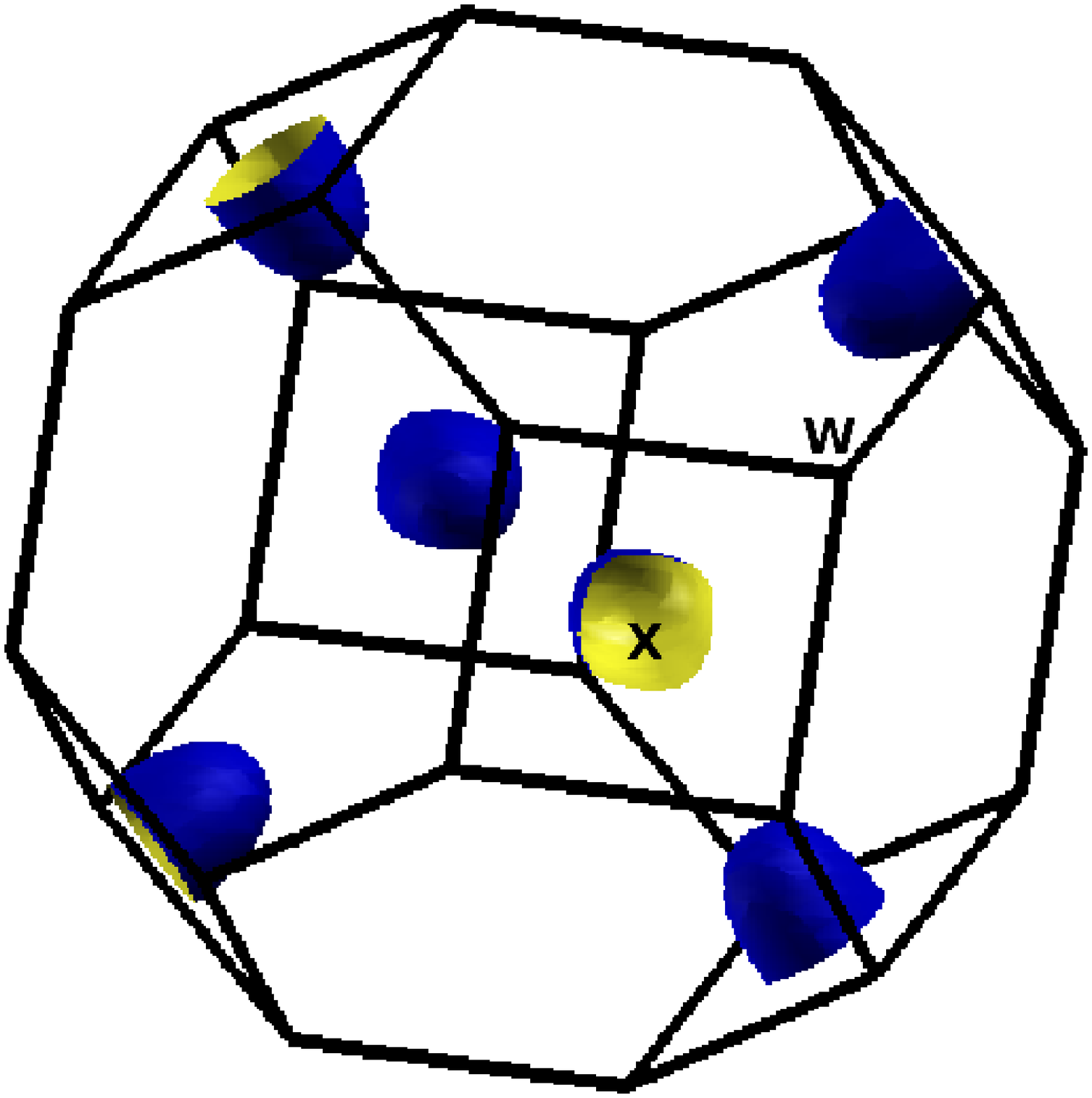}}\\
\subfigure[]{\includegraphics[width=35mm,height=35mm]{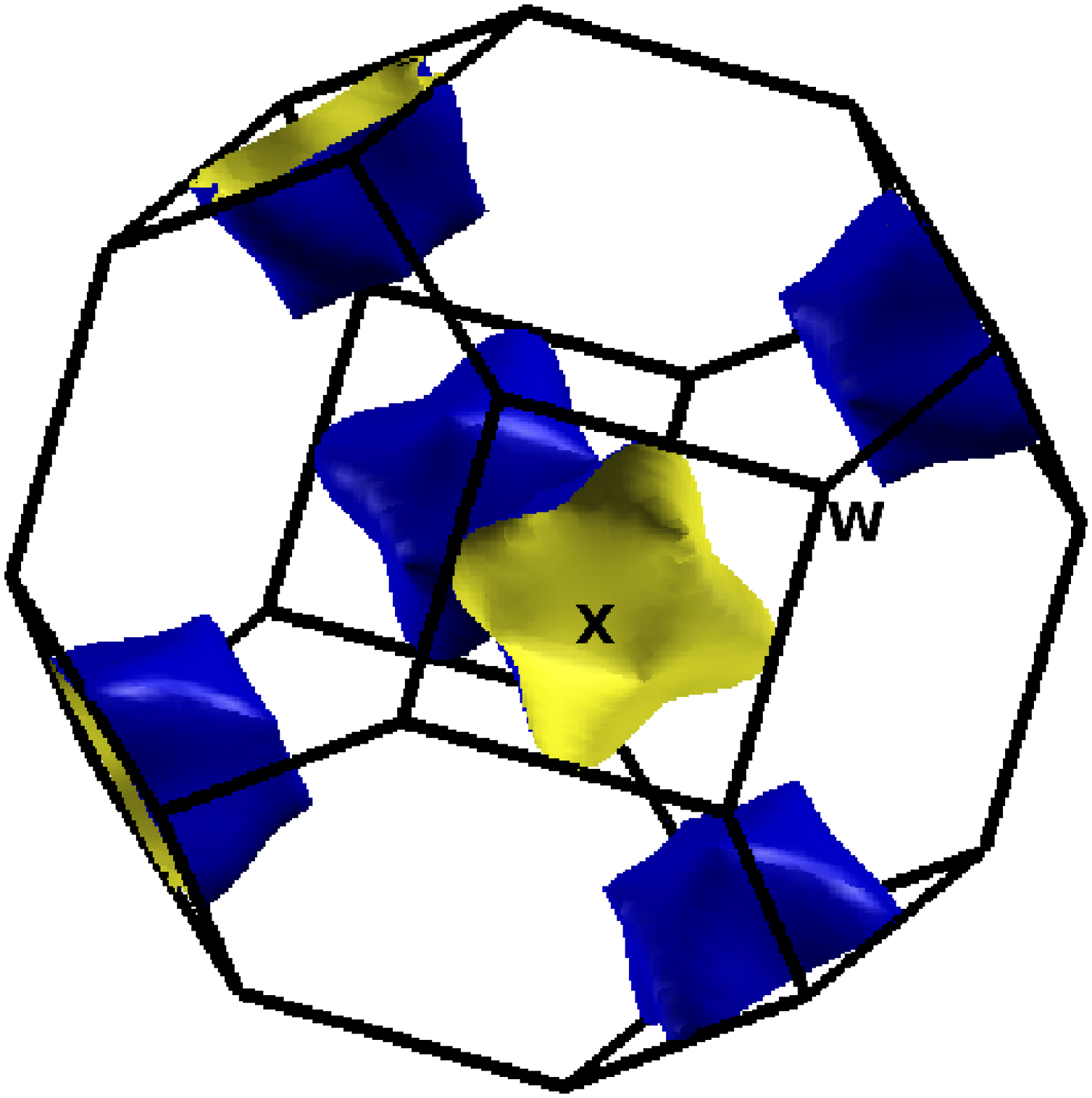}}
\subfigure[]{\includegraphics[width=35mm,height=35mm]{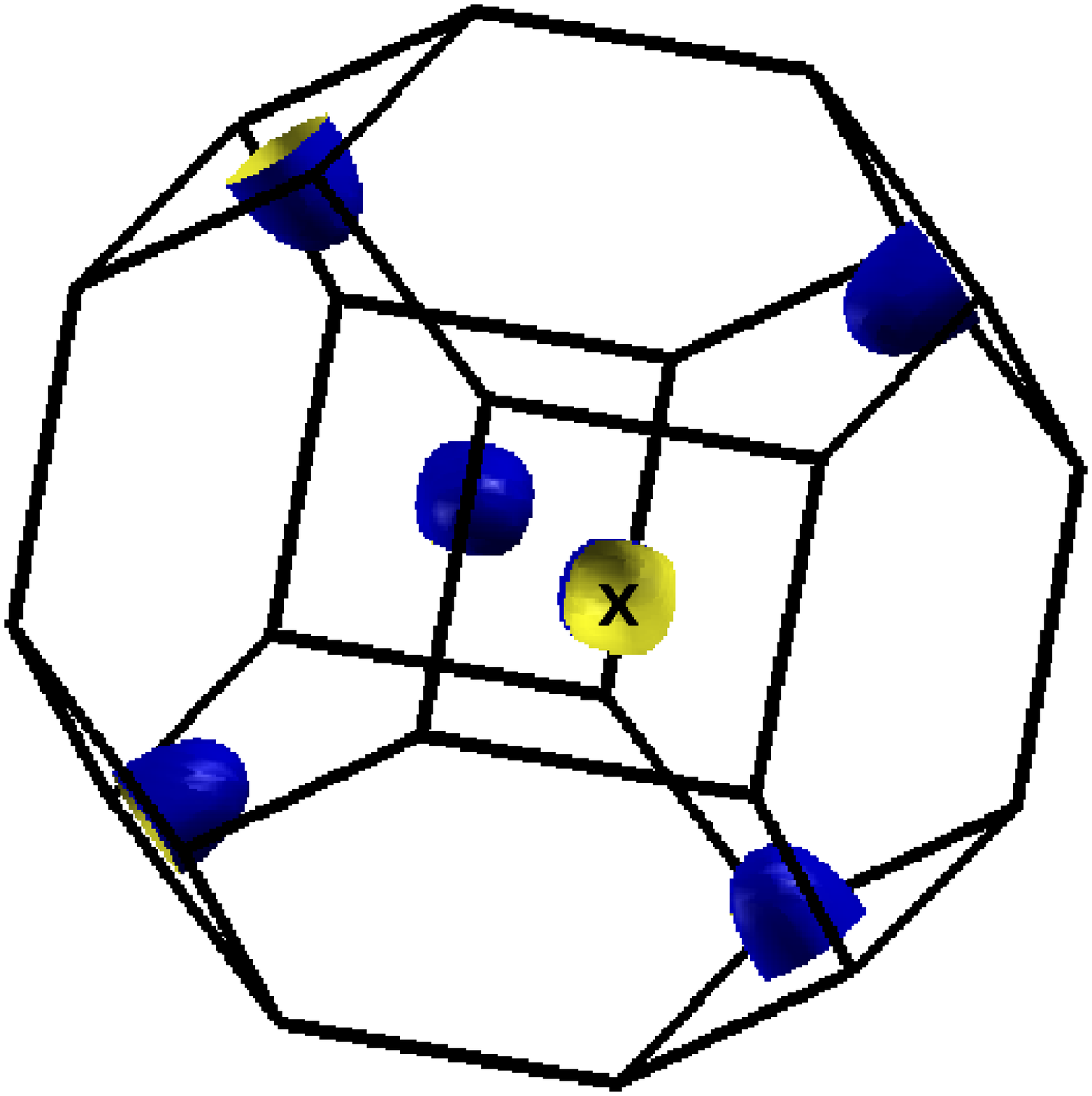}}
\caption{(colour online)Fermi surface for (a) Ni$_2$NbAl, (b) Ni$_2$NbGa, (c,d) Ni$_2$NbSn and (e,f) Ni$_2$VAl.}
\end{center}
\end{figure*}

\begin{figure*}
\begin{center}
\subfigure[]{\includegraphics[width=75mm,height=50mm]{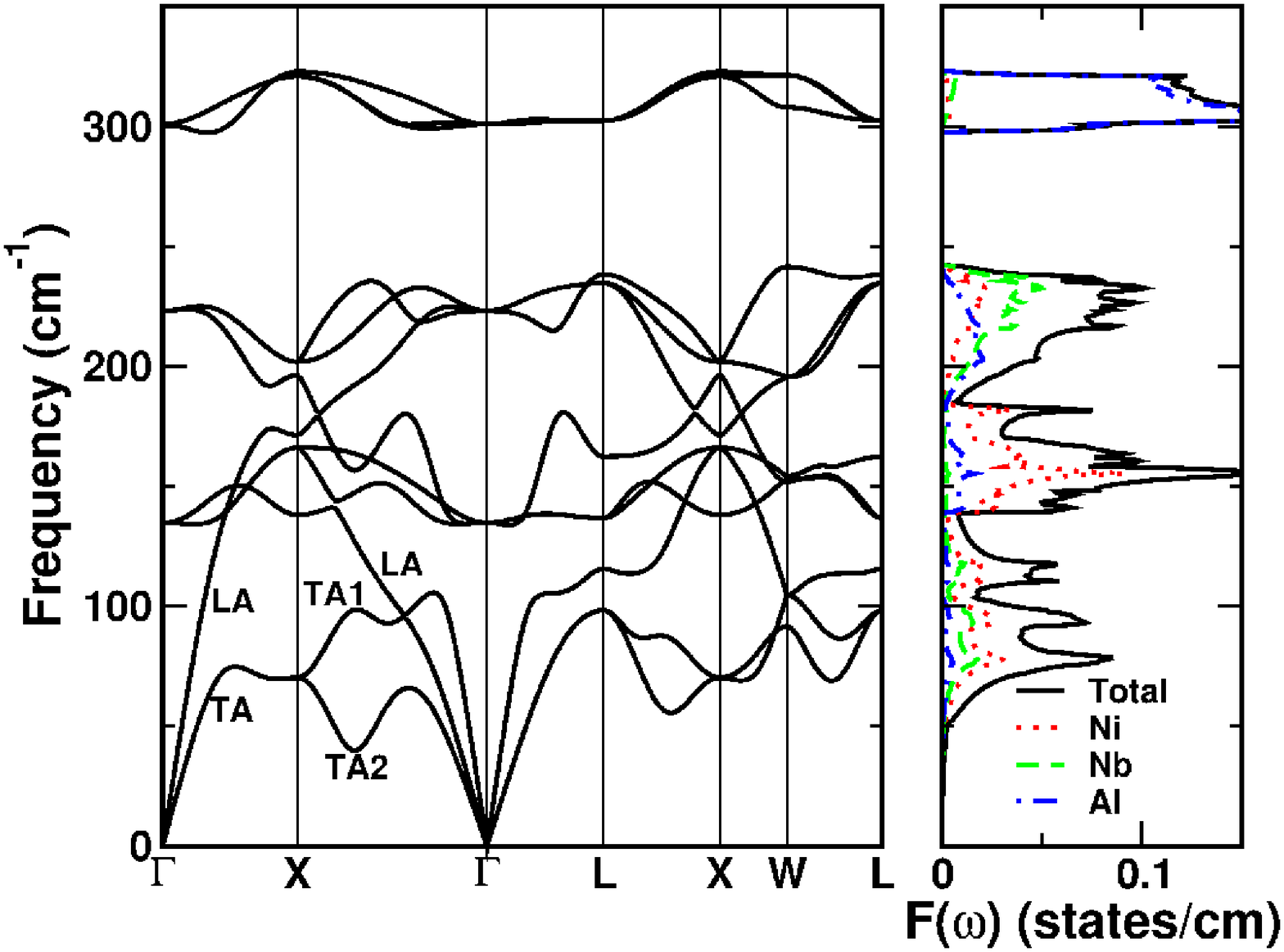}}
\subfigure[]{\includegraphics[width=75mm,height=50mm]{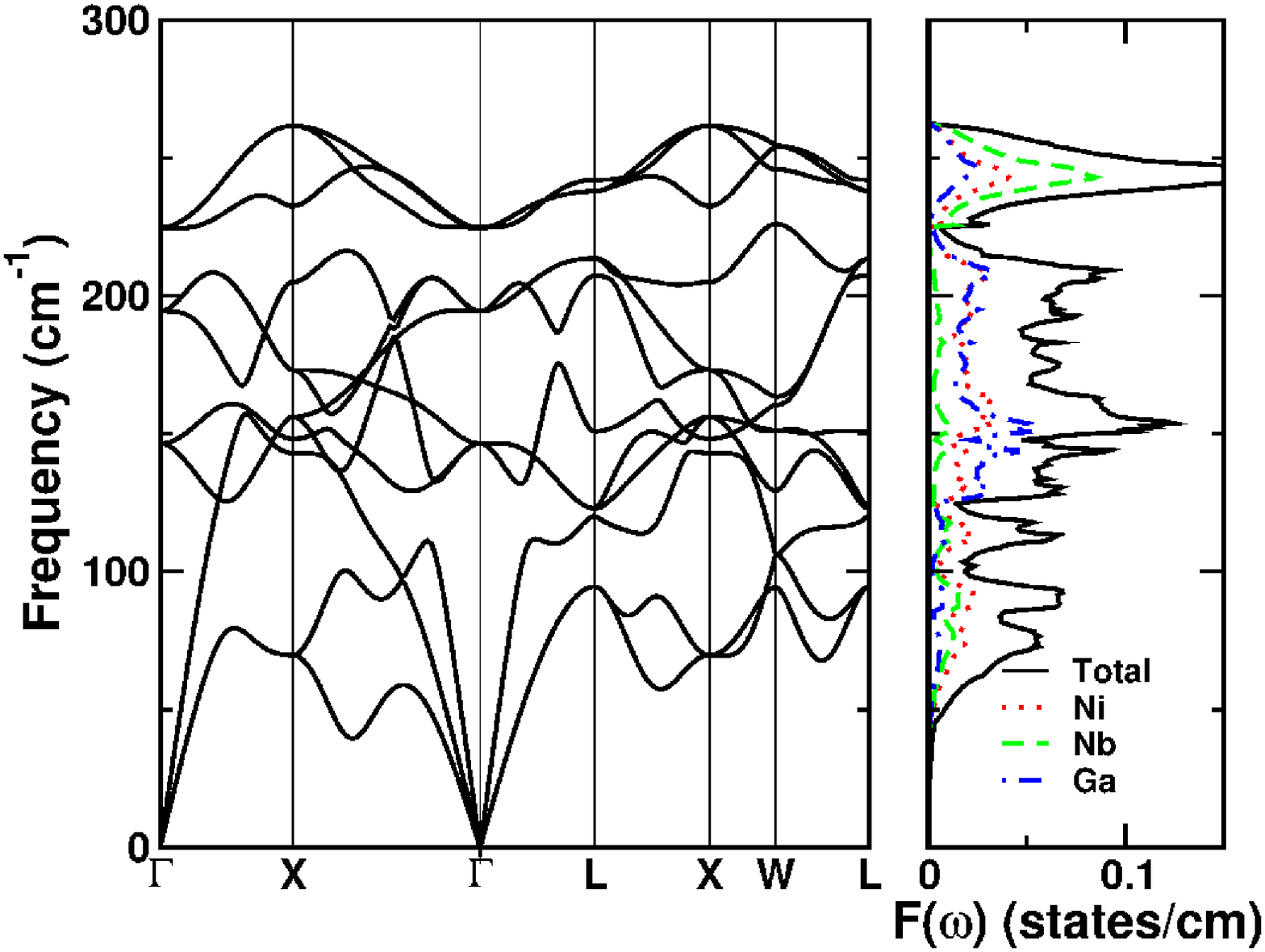}}
\subfigure[]{\includegraphics[width=75mm,height=50mm]{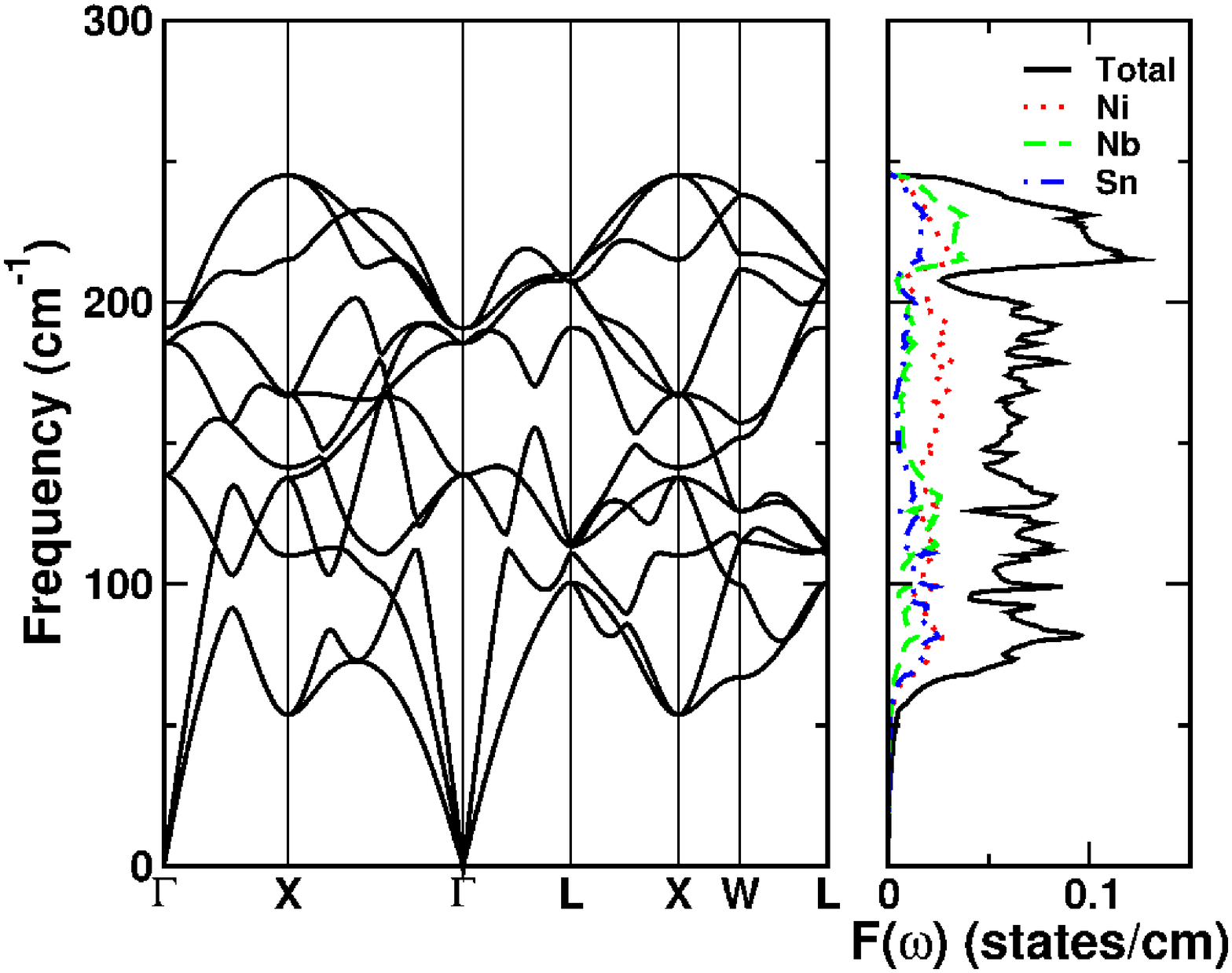}}
\subfigure[]{\includegraphics[width=75mm,height=50mm]{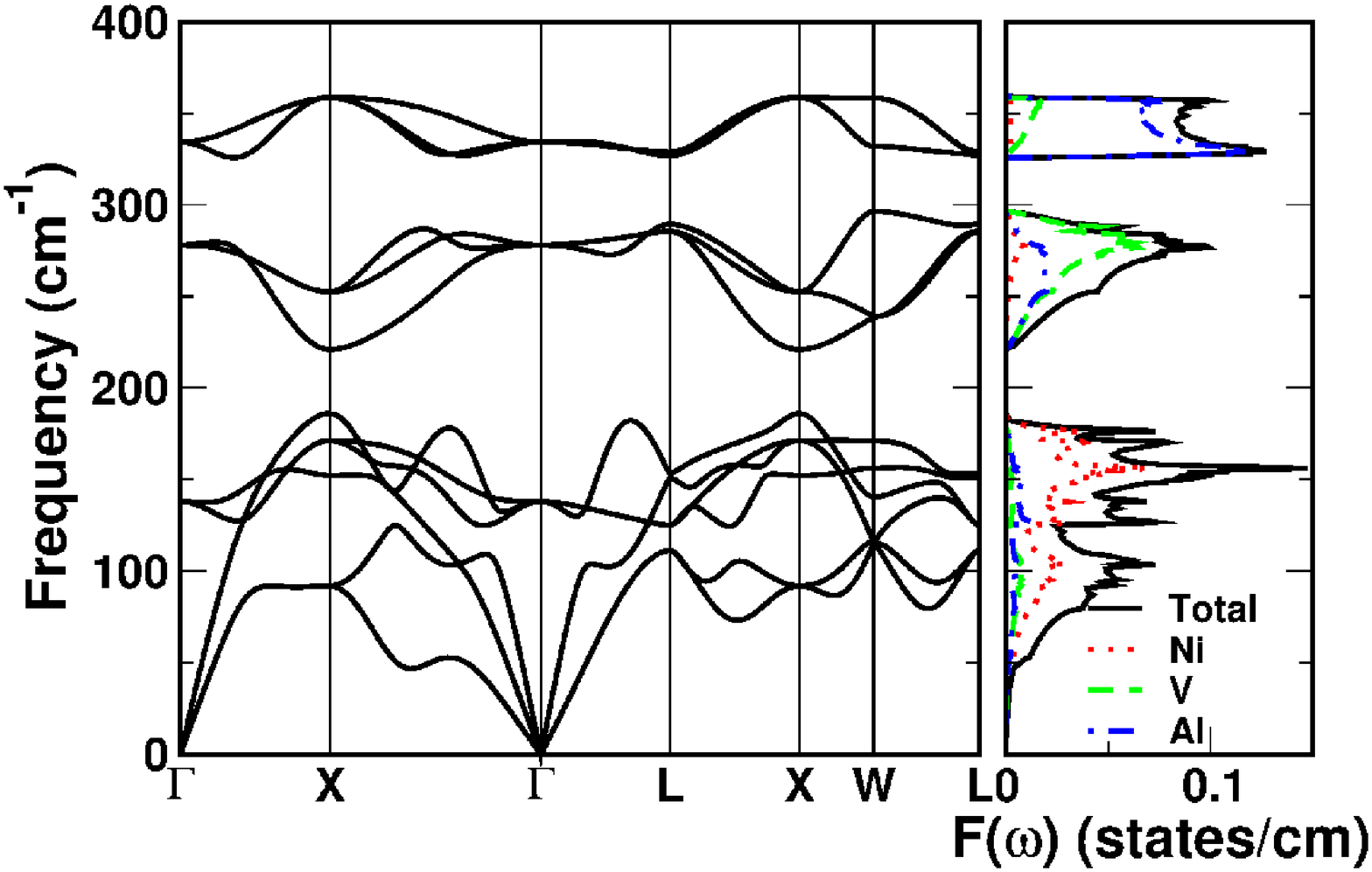}}
\caption{(colour online)Phonon dispersion along with partial phonon density of states  for (a) Ni$_2$NbAl, (b) Ni$_2$NbGa, (c) Ni$_2$NbSn and (d) Ni$_2$VAl.}
\end{center}
\end{figure*}

\begin{figure*}
\begin{center}
\subfigure[]{\includegraphics[width=50mm,height=50mm]{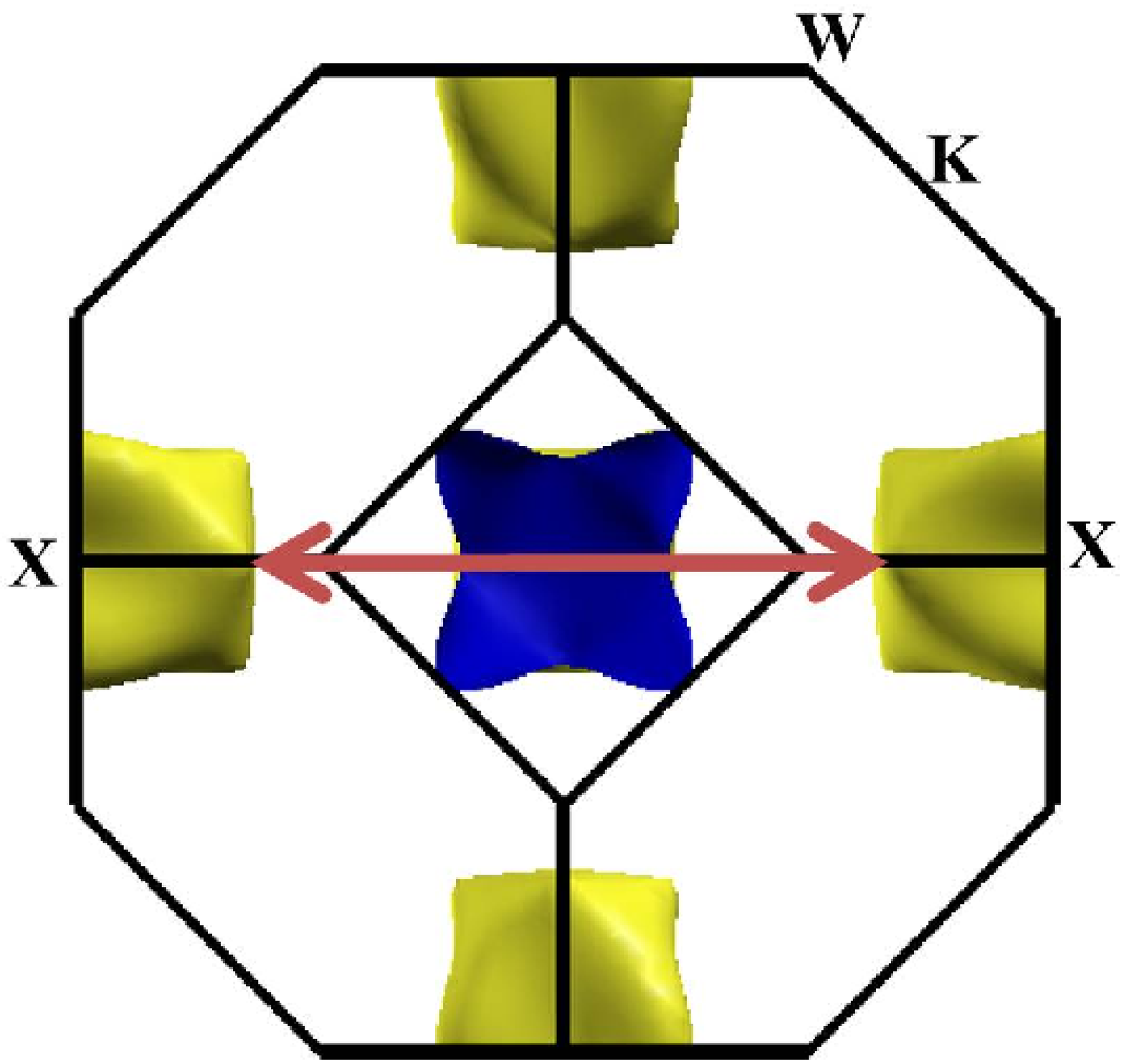}}
\subfigure[]{\includegraphics[width=60mm,height=50mm]{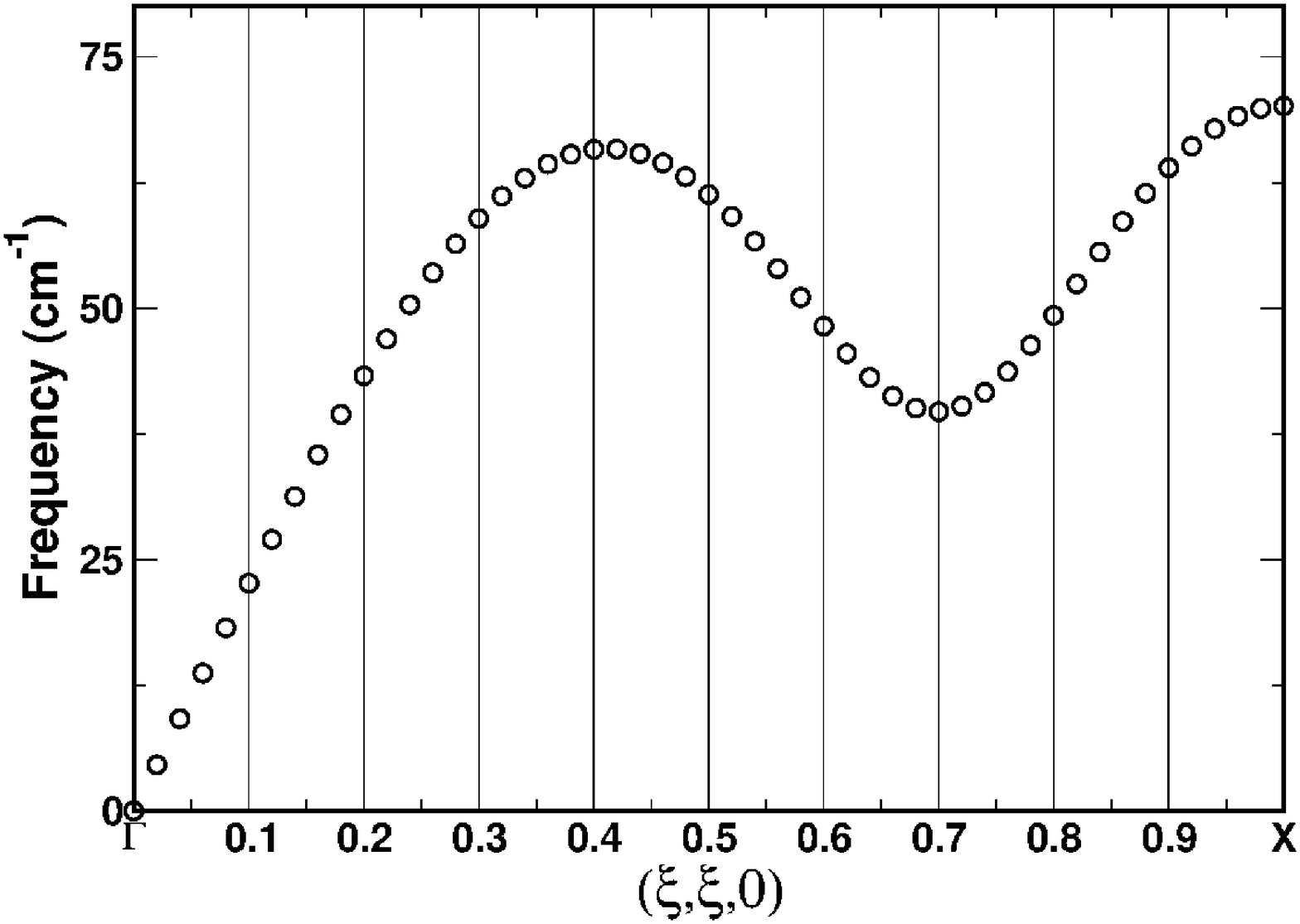}}\\
\subfigure[]{\includegraphics[width=60mm,height=50mm]{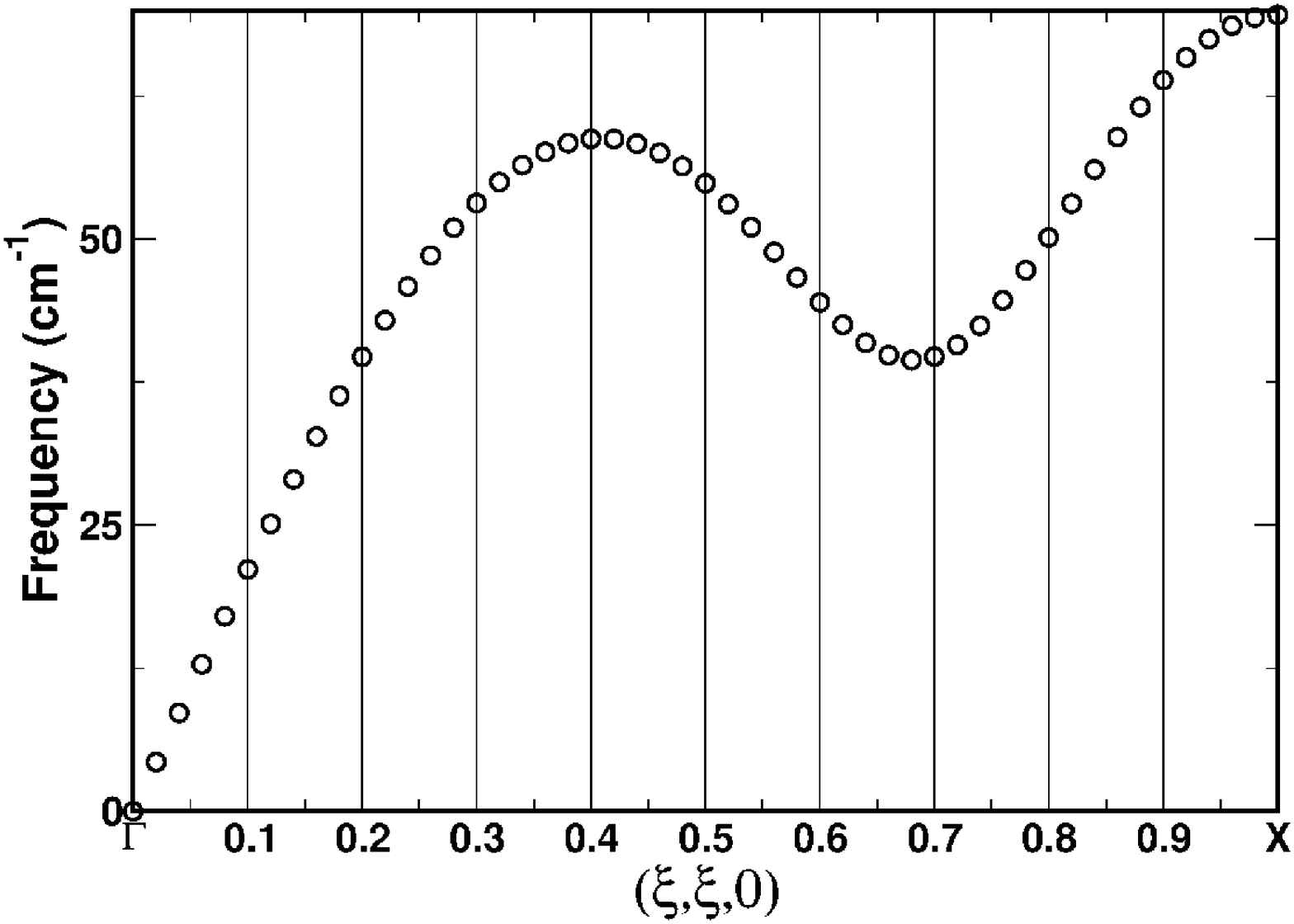}}
\subfigure[]{\includegraphics[width=60mm,height=50mm]{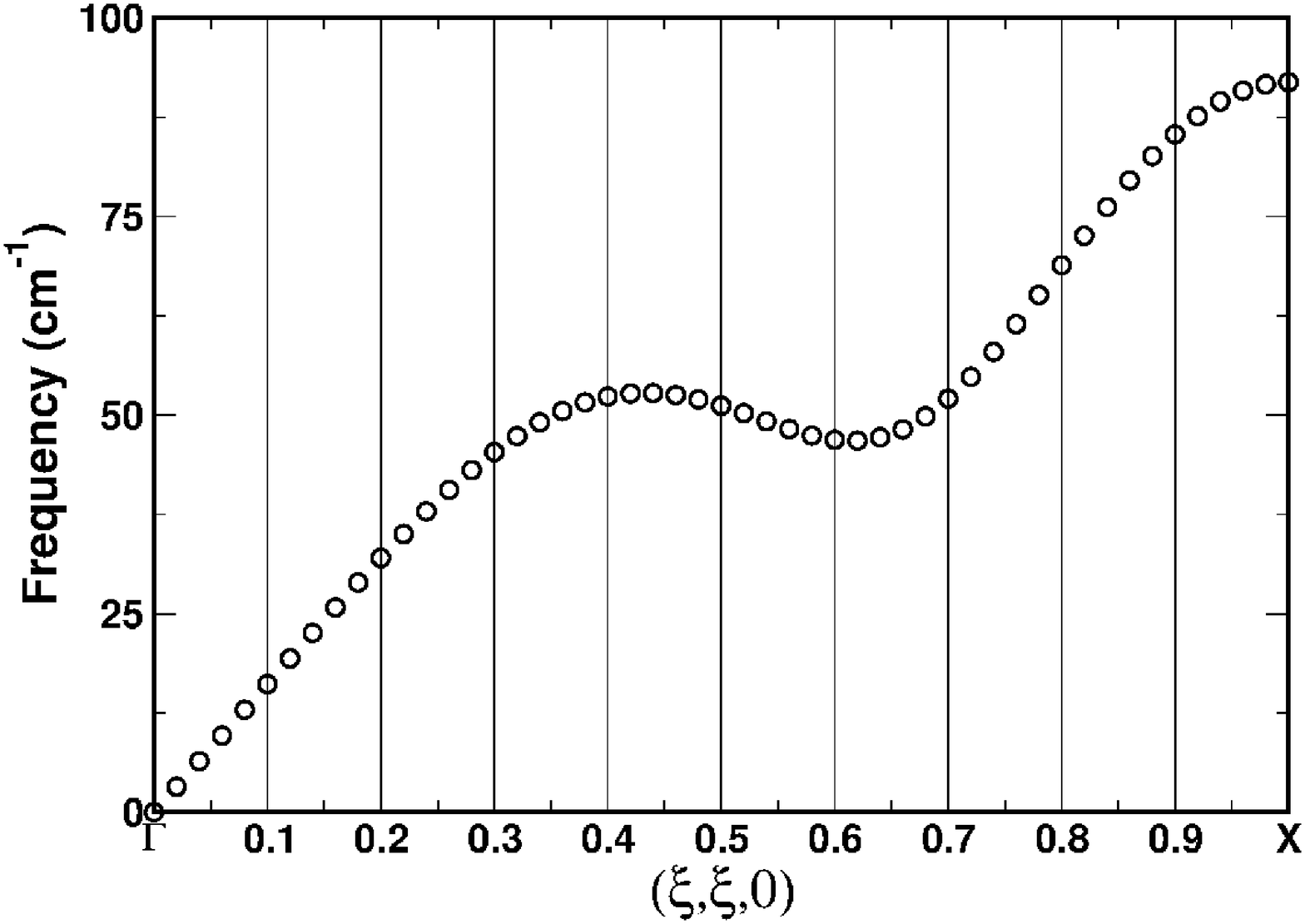}}
\caption{(colour online)(a) Direction of nesting vector and TA2 acoustic phonon mode along $\Gamma$-X direction in (b) Ni$_2$NbAl (c) Ni$_2$NbGa and (d) Ni$_2$VAl.}
\end{center}
\end{figure*}

\begin{figure*}
\begin{center}
\includegraphics[width=100mm,height=100mm]{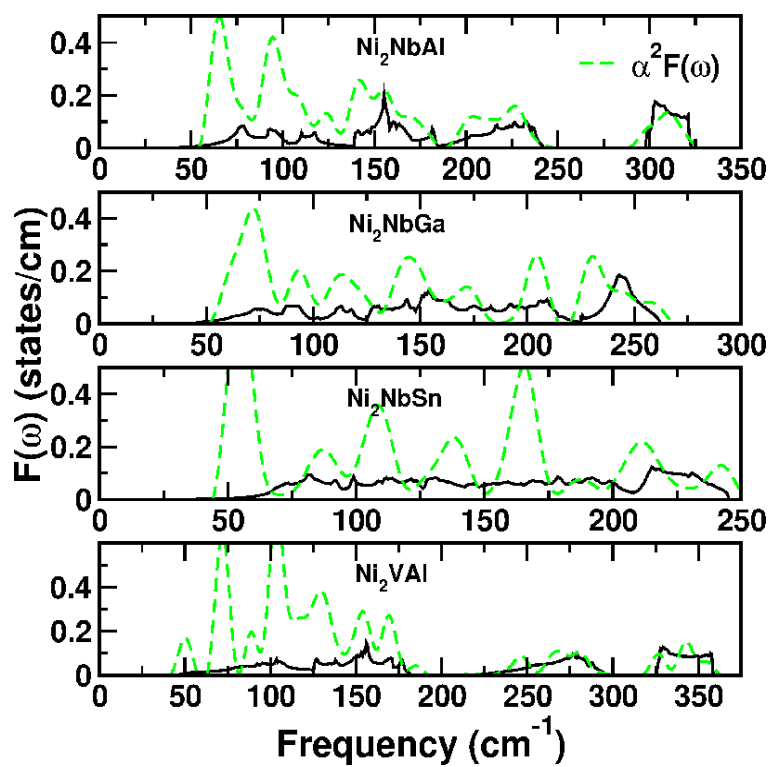}
\caption{(colour online)Eliasberg function $\alpha^2$F($\omega$) and phonon-density of states for for all compounds.}
\end{center}
\end{figure*}

\begin{figure*}
\begin{center}
\subfigure[]{\includegraphics[width=70mm,height=60mm]{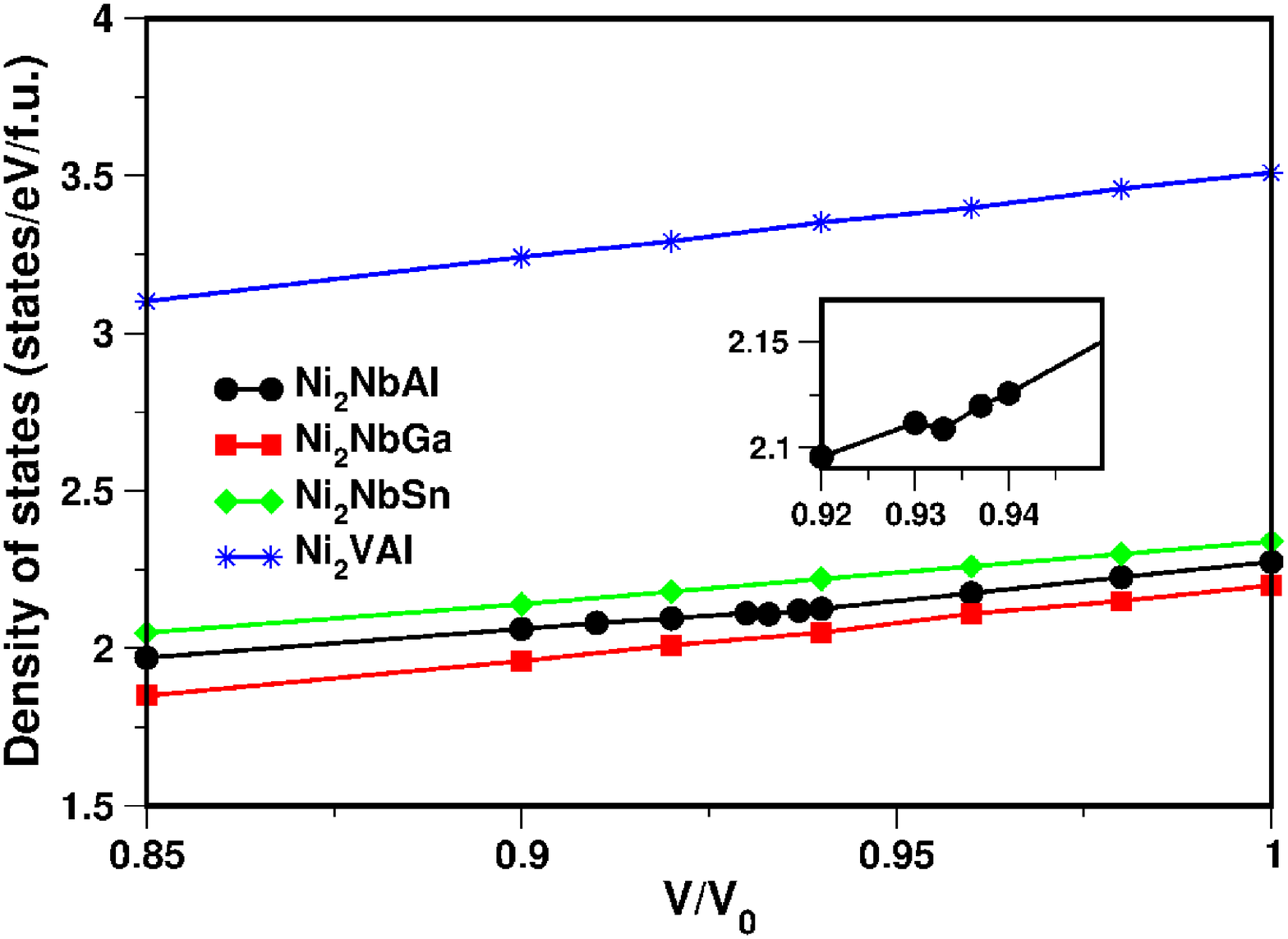}}
\subfigure[]{\includegraphics[width=70mm,height=60mm]{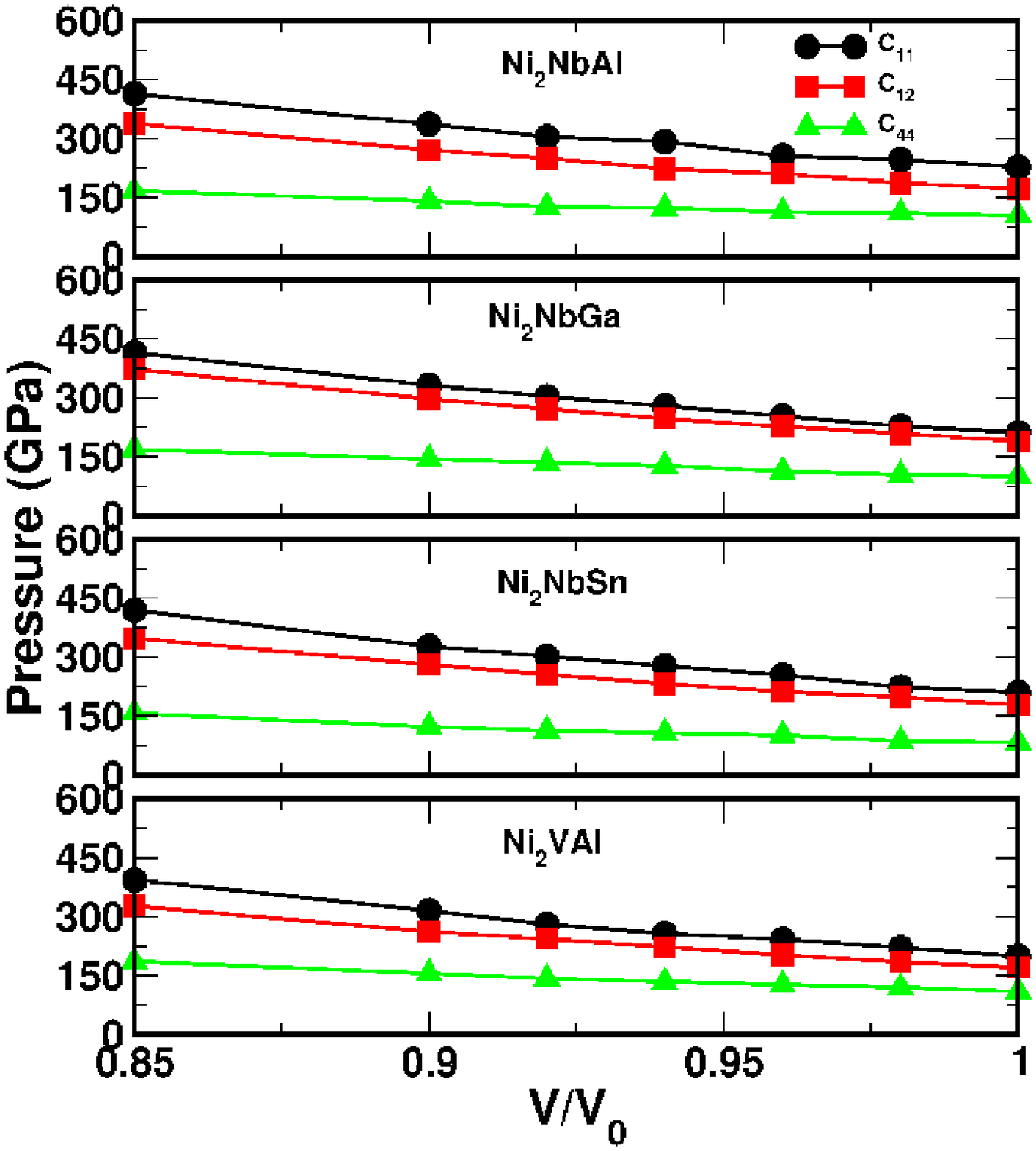}}
\caption{(colour online)(a) Electronic density of states under compression and (b) Elastic constants under compression for all compounds.}
\end{center}
\end{figure*}

\begin{figure*}
\begin{center}
\subfigure[]{\includegraphics[width=70mm,height=50mm]{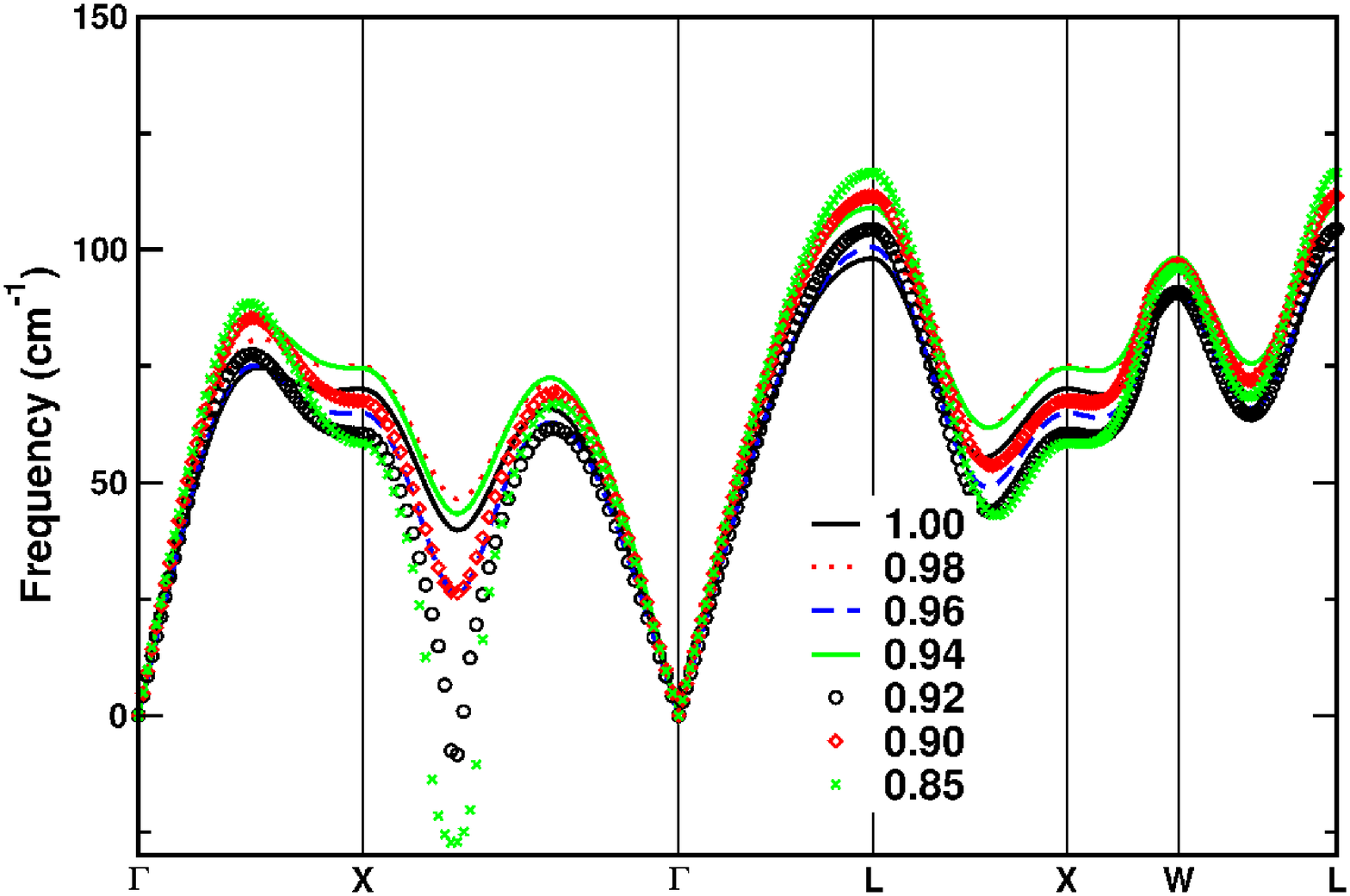}}
\subfigure[]{\includegraphics[width=70mm,height=50mm]{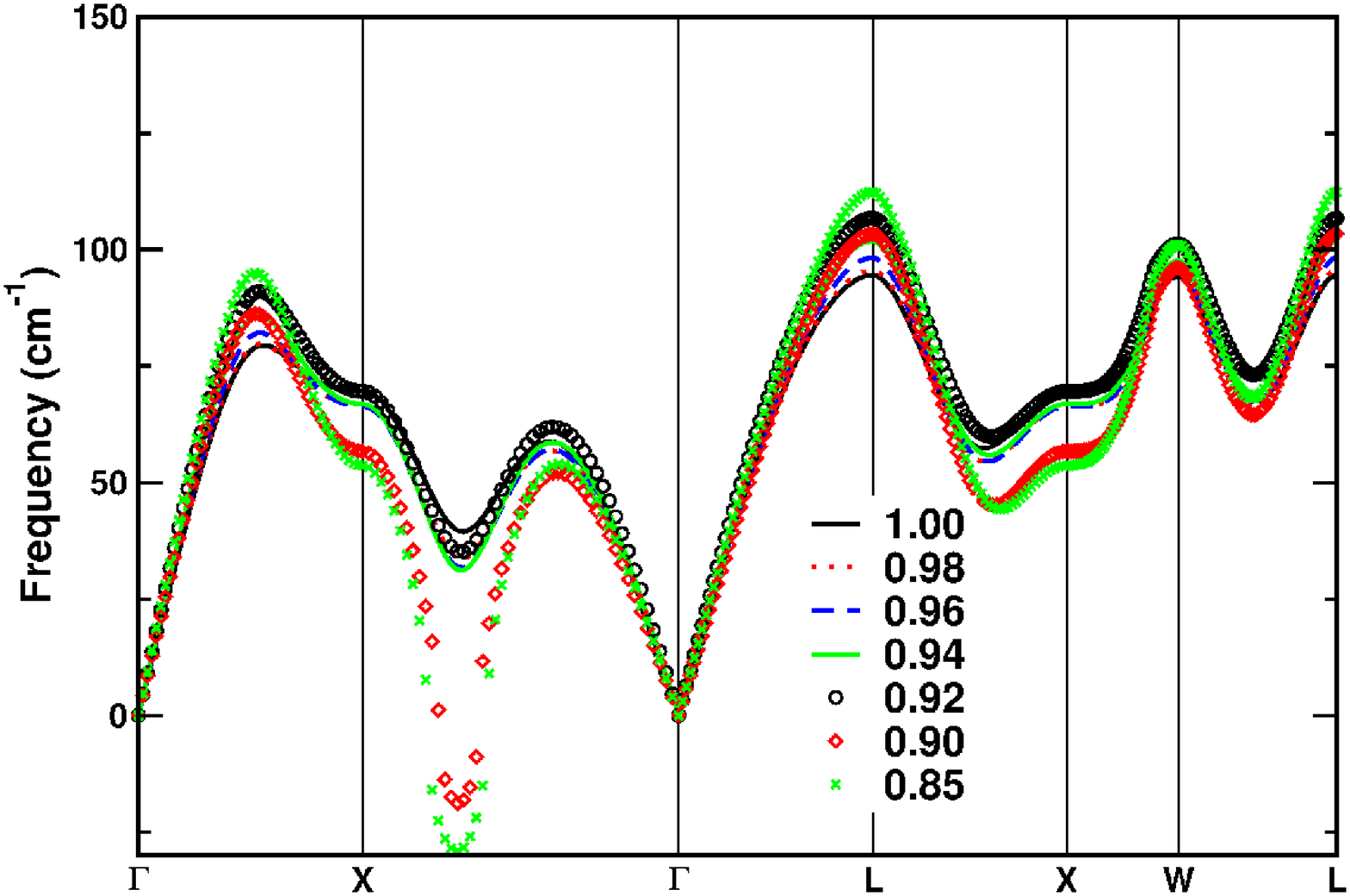}}
\subfigure[]{\includegraphics[width=70mm,height=50mm]{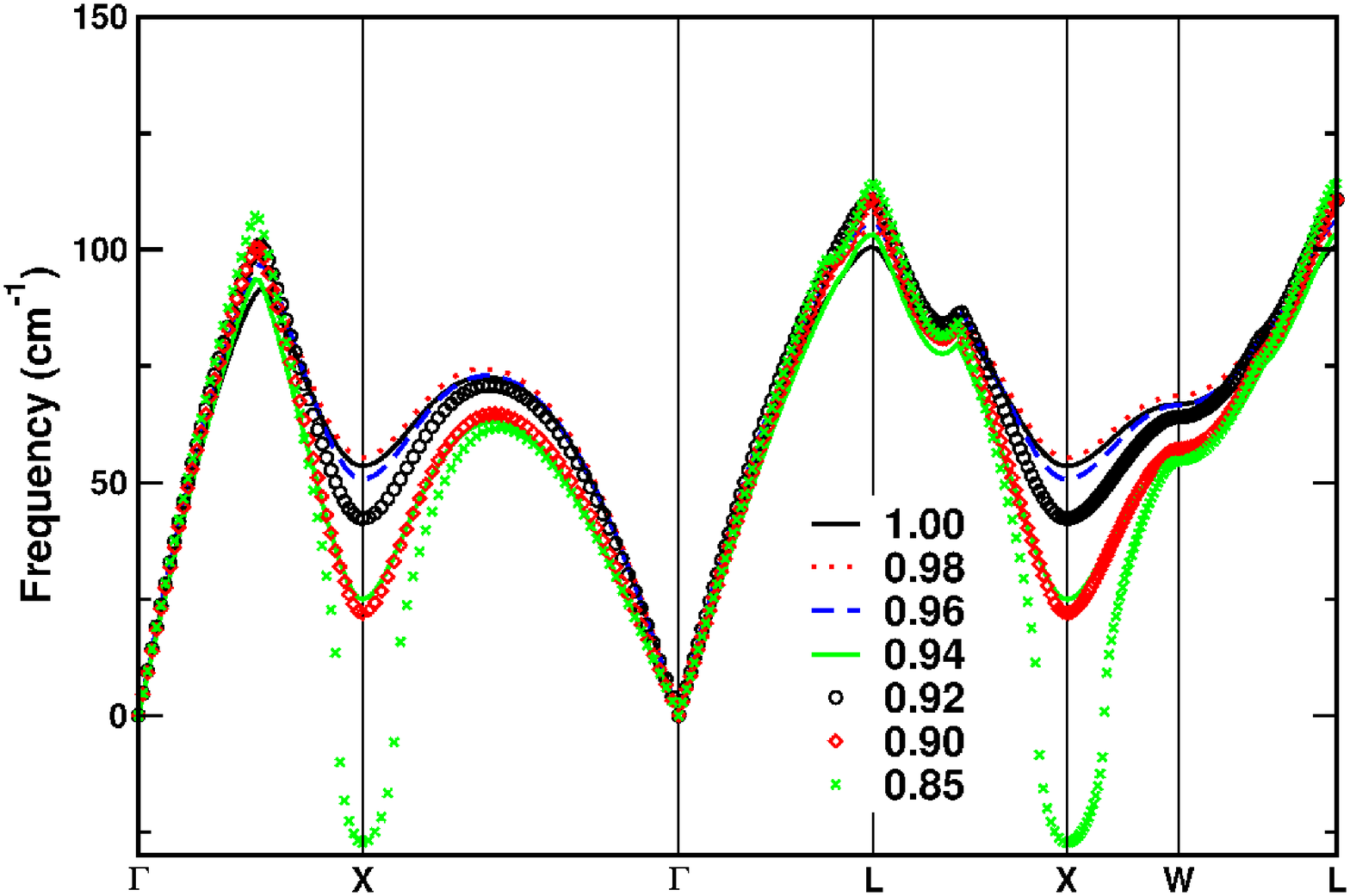}}
\subfigure[]{\includegraphics[width=70mm,height=50mm]{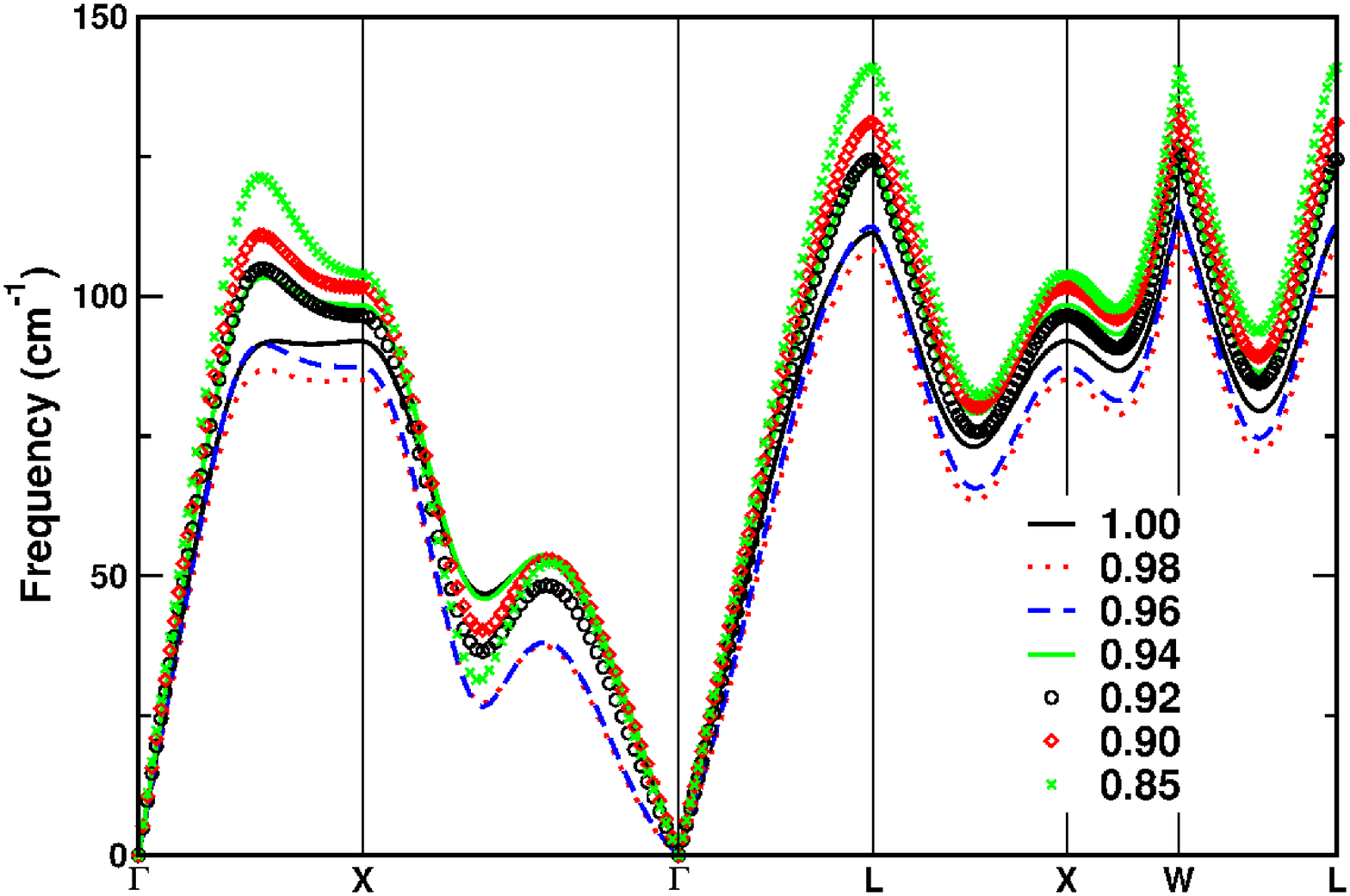}}
\caption{(colour online) Softening of lower frequency acoustic mode under compression in (a) Ni$_2$NbAl, (b) Ni$_2$NbGa, (c) Ni$_2$NbSn and (d)Ni$_2$VAl.}
\end{center}
\end{figure*}

\begin{figure}
\begin{center}
\includegraphics[width=150mm,height=80mm]{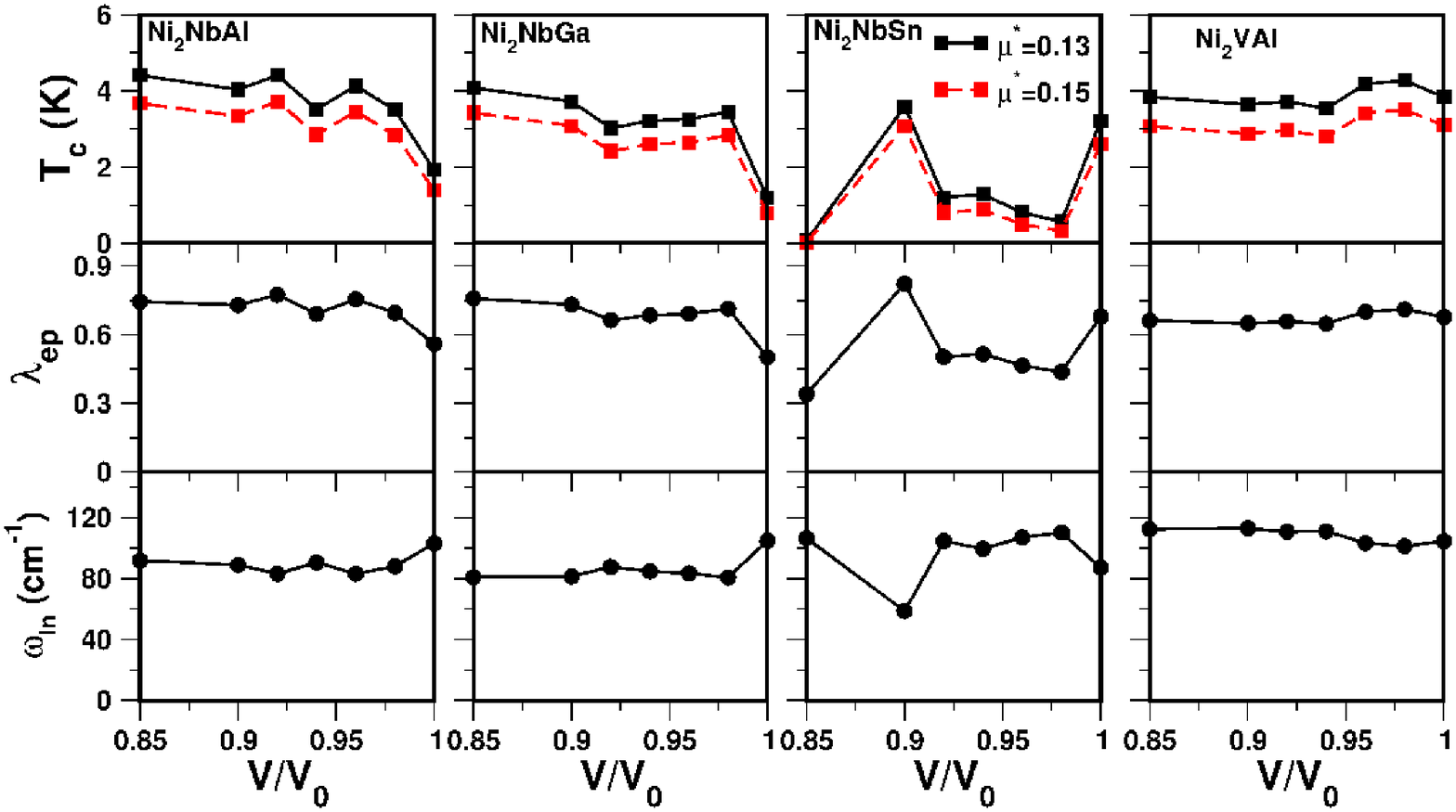}
\caption{(colour online)Logarithmic frequency, Electron-phonon coupling and superconducting transition temperature under compression for Ni$_2$NbX (X= Al, Ga and Sn) and Ni$_2$VAl compounds.}
\end{center}
\end{figure}

\end{document}